\documentclass[iop]{emulateapj}

\usepackage{subfigure}
\usepackage{multirow}

\newcommand{\spitzer}{\textit{Spitzer~}}   
\newcommand{\kms}{\textrm{km~s$^{-1}$}}
\newcommand{\lsolar}{L$_{\odot}$}
\newcommand{\msolar}{M$_{\odot}$}
\newcommand{\ml}{M$_{\odot}$ yr$^{-1}$}
\newcommand{\mdot}{\dot{M}} 
\newcommand{\radbbwarm}{$4.8\times10^{16}~{\rm cm~(0.048~ly)}$}
\newcommand{\radbbhot}{$7.7\times10^{15}~{\rm cm~(0.0077~ly)}$}

\begin{document}

\title{Disentangling the Origin and Heating Mechanism of Supernova Dust: Late-Time $Spitzer$~Spectroscopy of the Type IIn SN 2005ip}
\shorttitle{$Spitzer$~IRS Spectra of the Type IIn SN 2005ip}
\author{Ori D. Fox \altaffilmark{1,2}, Roger A. Chevalier \altaffilmark{1}, Eli Dwek \altaffilmark{2}, Michael F. Skrutskie \altaffilmark{1}, Ben E. K. Sugerman \altaffilmark{3}, Jarron M. Leisenring \altaffilmark{1}}
\altaffiltext{1}{Department of Astronomy, University of Virginia, Charlottesville, VA 22903}
\altaffiltext{2}{NASA Goddard Space Flight Center, Greenbelt, MD 20771}
\altaffiltext{3}{Department of Physics \& Astronomy, Goucher College, Baltimore, Maryland 21204}
\email{ofox@virginia.edu}

\begin{abstract}

This paper presents late-time near-infrared and {\it Spitzer} mid-infrared photometric and spectroscopic observations of warm dust in the Type IIn SN 2005ip in NGC 2906.  The spectra show evidence for two dust components with different temperatures.  Spanning the peak of the thermal emission, these observations provide strong constraints on the dust mass, temperature, and luminosity, which serve as critical diagnostics for disentangling the origin and heating mechanism of each component.  The results suggest the warmer dust has a mass of $\sim 5 \times 10^{-4}~$\msolar, originates from newly formed dust in the ejecta, or possibly the cool, dense shell, and is continuously heated by the circumstellar interaction.  By contrast, the cooler component likely originates from a circumstellar shock echo that forms from the heating of a large, pre-existing dust shell $\sim 0.01 - 0.05$~\msolar~by the late-time circumstellar interaction.  The progenitor wind velocity derived from the blue edge of the He 1 1.083 \micron~P Cygni profile indicates a progenitor eruption likely formed this dust shell $\sim$100 years prior to the supernova explosion, which is consistent with a Luminous Blue Variable (LBV) progenitor star.
\\
\\
\\
\end{abstract}


\newpage
\clearpage

\section{Introduction}
\label{sec3:intro}

SN 2005ip was discovered in NGC 2906 ($d \approx$ 30 Mpc) on UT 2005 November 5 \citep{boles05}.  Early optical spectra suggested the discovery occurred a few weeks following the explosion \citep{modjaz05}.  The supernova is a Type IIn event given the development of narrow H$\alpha$~emission lines \citep{smith09ip}.  \citet{fox09} discovered a late-time near-infrared excess that has persisted at least two and a half years post-discovery.

Late-time infrared emission typically indicates the presence of warm dust.  The origin and heating mechanism of the dust, however, is not always well constrained.  The dust may be newly formed or may have existed at the time of the supernova.  If newly formed, the dust may condense from refractory elements in the expanding supernova ejecta or in the cool, dense shell of post-shocked  circumstellar gas lying in between the forward and reverse shocks (like the post-shocked environment trailing stellar wind collisions in WR binary systems \citep{usov91}).  In both cases, several heating mechanisms are possible, including radioactivity, optical emission from circumstellar interaction, and collisional heating by hot gas in the reverse shock.  

Alternatively, pre-existing dust may be collisionally heated by hot, shocked gas or radiatively heated by either the peak supernova luminosity or late-time optical emission from circumstellar interaction.  In the latter case, an `IR echo' is evident due to light travel time effects.  Multiple scenarios can contribute to the late-time infrared flux, as in the cases of SNe 2004et \citep[][Sugerman et al. in prep]{kotak09} and 2006jc \citep{mattila08}.

Type IIn supernovae, named for their ``narrow'' emission lines \citep{schlegel90}, are more often observed to exhibit late time infrared emission associated with warm dust than any other supernova subclass \citep[e.g.,][]{pastorello02,gerardy02, pozzo04, fox09, smith09gy, miller10gy, miller10}.  Disentangling the origin and heating mechanism of the dust can yield important diagnostics concerning the circumstellar medium, supernova progenitor, and explosion dynamics.  For example, if the dust is shock heated, the observed dust temperature yields the gas density \citep{dwek87,dwek08}, which can be used to trace the progenitor's mass loss history \citep[e.g.,][]{smith09ip}.  If the dust is newly formed, the observed dust mass can be compared to models that predict supernovae as primary sources of dust at high redshifts \citep{todini01,nozawa03,nozawa08}.

\begin{deluxetable*}{c c c c c c c c}
\tablewidth{0pt}
\tablecaption{{\it Spitzer} Observations Summary \label{tab1}}
\tablecolumns{8}
\tablehead{
\colhead{Target} & \colhead{Position} & \colhead{Flux} & \colhead{} &
\colhead{Int/} & \colhead{Total} & \colhead{Ramp $\times$} & \colhead{\# of}\\
\colhead{Field} & \colhead{RA} & \colhead{Density} & \colhead{Bands} & \colhead{Pixel} & \colhead{Duration} & \colhead{Cycles} & \colhead{AORS}\\
\colhead{} & \colhead{DEC} & \colhead{} & \colhead{} & \colhead{(secs)} & \colhead{(secs)} & \colhead{(sec $\times$ \#)} & \colhead{}
}
\startdata
 SN 2005ip & 09:32:06 & 0.4 mJy & IRAC & 300 & 1062 & - &  1 \\
& +08:26:44& (8\micron)& all& & &  &\\
& & & & &  & &\\
 SN 2005ip &  09:32:06 &  0.4 mJy & IRS & - & 9516 & 60 $\times$ 12 &  1 \\
&  +08:26:44& (8\micron) & Spectral Mapping & & & & \\
& & & low short both & & &  &\\
\enddata
\end{deluxetable*}

The origin of the observed dust is not always obvious.  For SN 2005ip, optical spectra show a progressive attenuation of the red wing of both the broad and intermediate lines, directly confirming the formation of new dust in both the ejecta and post-shocked cool, dense shell \citep{smith09ip}.  Such direct evidence, however, is rare.  No more than a handful of supernovae \citep{lucy91,meikle93,elmhamdi04,elmhamdi03,sugerman06,kotak08,pozzo04,smith09ip,smith08jc} show direct evidence of dust formation in the ejecta, and aside from SN 2005ip, only SNe 1998S \citep{pozzo04} and 2006jc \citep{smith08jc} show direct evidence of dust formation in the cool, dense shell (although this scenario is invoked to explain the asymmetries observed in the spectra of SNe 2004et \citep{kotak09} and 2007od  \citep{andrews10}).  The observed dust yields, however, all tend to be 2-3 orders of magnitude lower than required to account for the large amounts of dust observed at high redshifts.

Mid-infrared ($3 \le \lambda \le 15$~\micron) observations span the peak of the thermal spectral energy distribution from dust with temperatures ranging $100 \lesssim T_d \lesssim 1000$~K, providing strong constraints on the dust mass, temperature, and, thereby, the luminosity.  These quantities serve as useful diagnostics for disentangling the origin and heating mechanism of warm dust.  Late-time mid-infrared observations of supernovae, particularly the Type IIn subclass, are rare.

In this paper, we present the first late-time (day 936 post-discovery) $Spitzer$/IRS spectra of a Type IIn supernova, as well as coincident $Spitzer$/IRAC photometry.  We also present 0.9-2.5 \micron, $R=3000$~spectra obtained with APO 3.5-m/TripleSpec.  Section \ref{sec3:obs} presents the observations and data reduction techniques. The combined spectra show evidence for two independent dust components: a hot, near-infrared (HNI) and warm, mid-infrared (WMI) component.  For each component, we derive the dust composition, mass, temperature, and luminosity.  Section \ref{sec3:source} explores the origin and heating mechanism of these components to determine the degree to which SN 2005ip forms new dust.  The HNI dust mass originates predominantly from newly formed dust in the ejecta, while the WMI component likely originates from an `IR echo' in a pre-existing dust shell.  We use these results to explore the progenitor system and its evolution.  Section \ref{sec3:con} presents a summary of the findings and a discussion of future work.

\section{Observations}
\label{sec3:obs}

\subsection{\spitzer}
\label{sec3:spitz}

As part of PID 50256, the $Spitzer$~Infrared Spectrograph (IRS) \citep{houck04} obtained one mid-infrared spectra on June, 3 2008 (936 days post-discovery) with the Short-Low module (SL, R$\sim$60--120, 5.2-14 \micron).  The $Spitzer$~Infrared Array Camera (IRAC) \citep{fazio04} followed the IRS observations with images of the supernova and host galaxy in all four bands on June 10, 2008.  Table \ref{tab1} lists the observational details.

\begin{figure}[t]
\epsscale{0.9}
\plotone{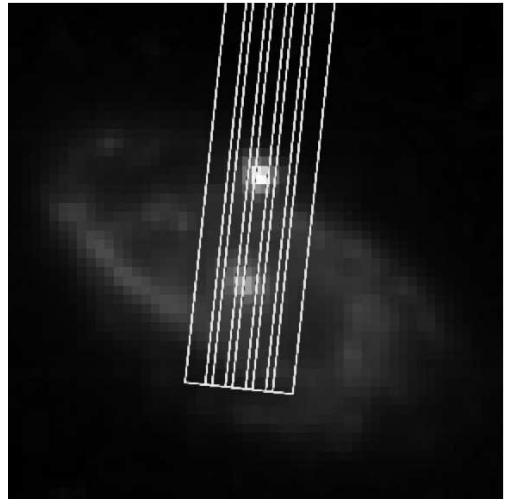}
\caption{{\it Spitzer}/IRAC and IRS observations of SN 2005ip in NGC 2906 obtained in June, 2008 ($\sim$943 and 936 days post-discovery, respectively).  Shown here is the Post-BCD 3.5 \micron~IRAC image with the IRS map overlay.  SN 2005ip is directly above the galaxy, while the galaxy nucleus also falls within the mapping scheme.}
\label{f1}
\end{figure}

\subsubsection{IRAC}

$Spitzer$~collected 5-minute IRAC exposures consisting of ten 30-second integrations.  Pipeline-reduced and calibrated (BCD) images were taken from the {\em Spitzer} archive, and combined into single frames with enhanced pixel resolution of 0\farcs75~pix$^{-1}$ using the {\tt MOPEX} software package provided by the Spitzer Science Center.  Figure \ref{f1} shows a post-BCD 3.5 \micron~IRAC image with the IRS map overlay.  Since the supernova lies near the center of the NGC 2906, the rapidly-varying background of the host galaxy complicates aperture photometry.  Instead, a number of unsaturated, linear, and isolated (e.g. no other sources in the wings) stars were used to build an empirical PSF for each channel, each of which was used to measure the brightness of the supernova.  In each channel, the residuals from subtracting the best-fit PSF were small compared with the predicted uncertainty that the photometry task {\tt allstar} provides, which factors in Poisson noise along with flat-field and profile-fitting errors as well as read noise.  PSF-fit measurements of field stars in the frame were consistent with aperture photometry to within 5\% in all channels.

\subsubsection{IRS}

\begin{figure}[t]
\epsscale{1.2}
\plotone{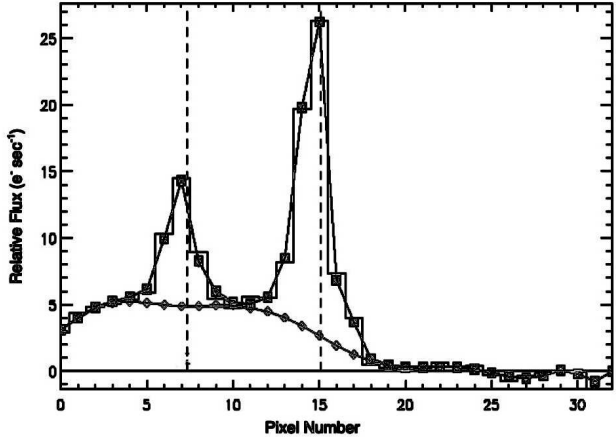}
\caption{Spatial profile showing SN 2005ip centered in the slit with the galaxy nucleus at the left shown by the {\tt AdOpt} tool within the {\tt SMART} data analysis package.  The two point sources are fit simultaneously along with the underlying background. }
\label{f2}
\end{figure}

The $Spitzer$/IRS mapped the position of SN 2005ip, with 12 cycles of 5 explosures, stepped 2\farcs7 perpendicular to the 3\farcs7 wide slit.  The target spectrum was extracted and calibrated from the central pointing observation using the {\tt SMART} data analysis software \citep{higdon04}, with bad pixels identified by {\tt IRSCLEAN}.  Subtracting off-order observations removed sky and zodiacal background.  All data collection events for a given order were then median combined and the spectra extracted using the advanced optimal extraction routine {\tt AdOpt} \citep{lebouteiller09}.  The {\tt AdOpt} tool provides the powerful ability to simultaneously fit multiple sources using a super-sampled PSF plus a complex background at each row.  This feature disentangles the supernova emission from that of the galaxy nucleus as well as removes background emission of the galaxy's spiral arms (Figure \ref{f2}).

Figure \ref{f3} shows the resulting IRS spectrum redshifted to account for the radial velocity of the galaxy (2140 \kms + V$_{LSR}$), along with the IRAC photometry.  Continuum emission peaking at around 3-4 \micron~tends to dominate the spectrum.  Few, if any, emission lines are apparent.  

\subsection{TripleSpec}
\label{sec3:tspec}

TripleSpec, an 0.9-2.5 \micron, $R=3000$~spectrograph operating at APO \citep{wilson04,herter08} obtained a spectrum on day 862 post-discovery.  Forty minutes of on-source integration consisted of 8 independent 5 minute exposures nodding between 2 different slit positions.  We extract the spectra with a modified version of the IDL-based {\it SpexTool} \citep{cushing04}.  The underlying galactic arm and sky are approximated in {\it SpexTool} by a polynomial fit and subtracted from the supernova.  TripleSpec observations on day 1243 post-discovery suggest little evolution occurred between the two epochs.  We therefore create a single spectrum from the near- and mid-infrared spectra (see Figure \ref{f3}).

\subsection{Dust Composition, Temperature, and Mass}
\label{sec3:dust}

Assuming only thermal emission, the combined near- and mid-infrared spectra provide a strong constraint on the dust mass, temperature, and thereby, the luminosity.  The luminosity of a single spherical dust particle of radius, $a$, and temperature, $T_d$, is given as
\begin{equation}
\label{eqn:lumparticle}
L_d (\lambda) = 4 \pi a^2 (\pi B_\nu(T_d) Q_\nu(a)),
\end{equation}
where $B_\nu(T_d)$~is the Planck blackbody function and $Q_\nu(a)$ is the emission efficiency.  \citet{fox09} and \citet{smith09ip} provide two pieces of evidence for optically thin dust: 1) $\tau \approx \frac{L_{\rm NIR}}{L_{\rm NIR} + L_{\rm OPT}} \approx 0.5$ and 2) the relatively high transmission of X-rays responsible for ionizing the unshocked circumstellar medium.  For optically thin dust with mass, $M_d$, at a distance, $d$, from the observer, thermally emitting at a single equilibrium temperature, the total flux can be written as:
\begin{equation}
\label{eqn:flux2}
F_\nu = \frac{M_{d} B_\nu(T_d) \kappa_\nu(a)}{d^2},
\end{equation}
where $\kappa_\nu(a)$, the dust mass absorption coefficient, is:
\begin{equation}
\label{eqn:kappa}
\kappa_\nu (a) = \Big(\frac{3}{4 \pi \rho a^3}\Big) (\pi a^2 Q_\nu(a)),
\end{equation}
for a dust bulk (volume) density $\rho$.  

\begin{figure}
\begin{center}
\epsscale{1.2}
\plotone{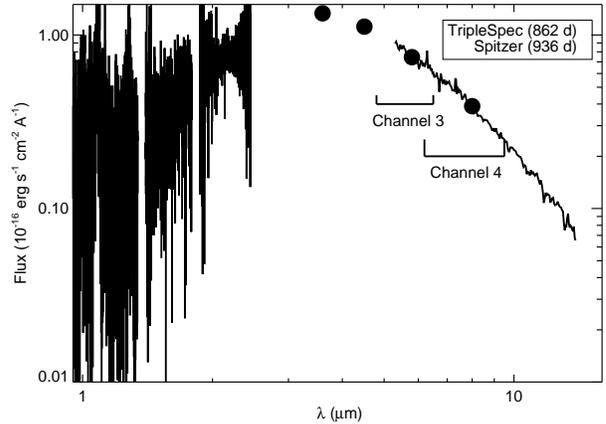}
\caption{Combined APO/TripleSpec near-infrared spectra and {\it Spitzer}/IRAC and IRS mid-infrared data.  Although the epochs do not match perfectly, TripleSpec observations on day 1243 post-discovery (not shown) suggest little evolution occurred since the day 862 spectrum plotted here.  We therefore treat the combined spectra as a single spectrum.  The thermal emission that dominates the spectrum confirms the presence of warm dust.}
\label{f3}
\end{center}
\end{figure}

\begin{figure*}
\begin{center}
\subfigure{
\label{f4a}
\epsscale{0.5}
\plotone{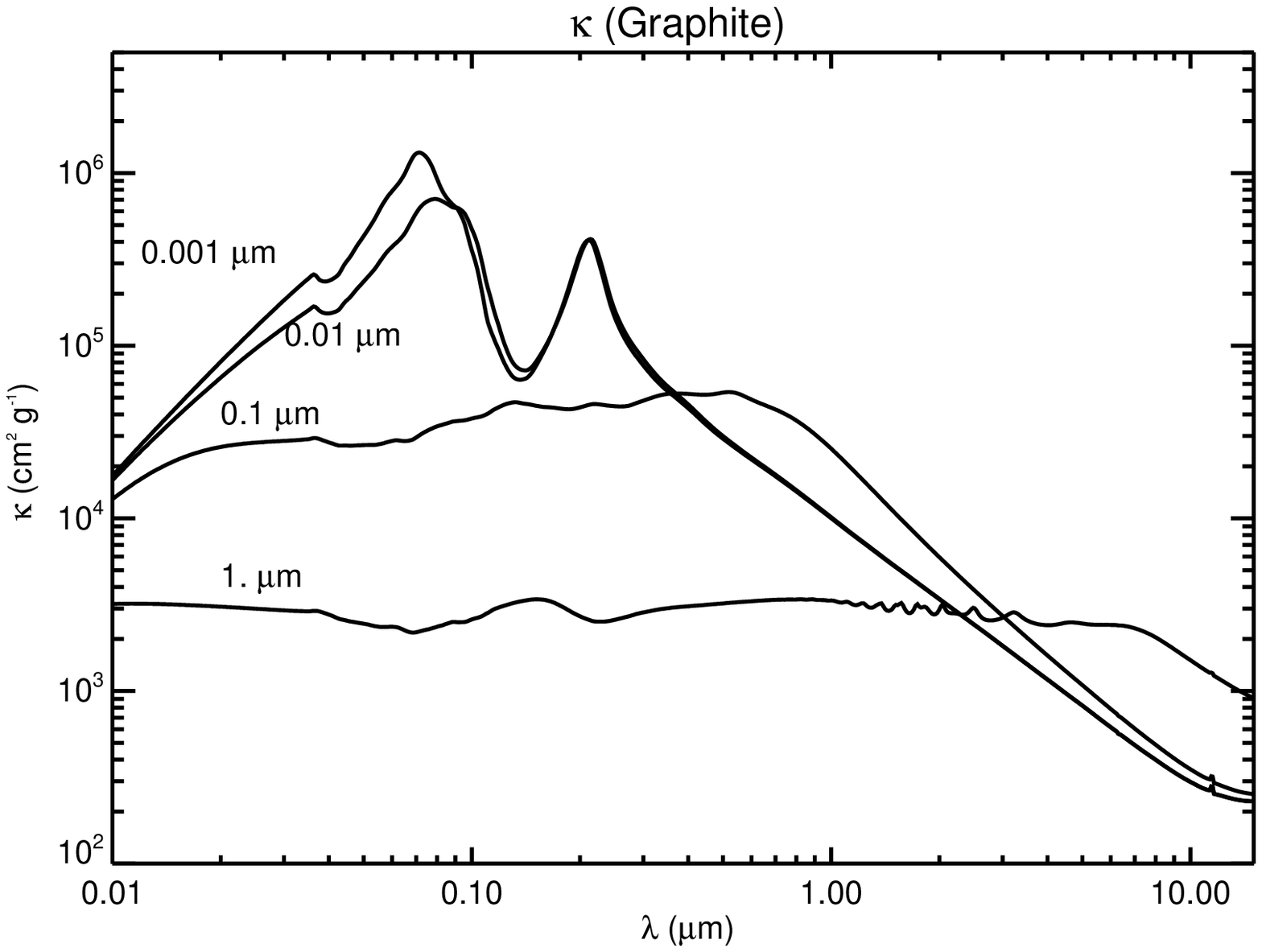}\plotone{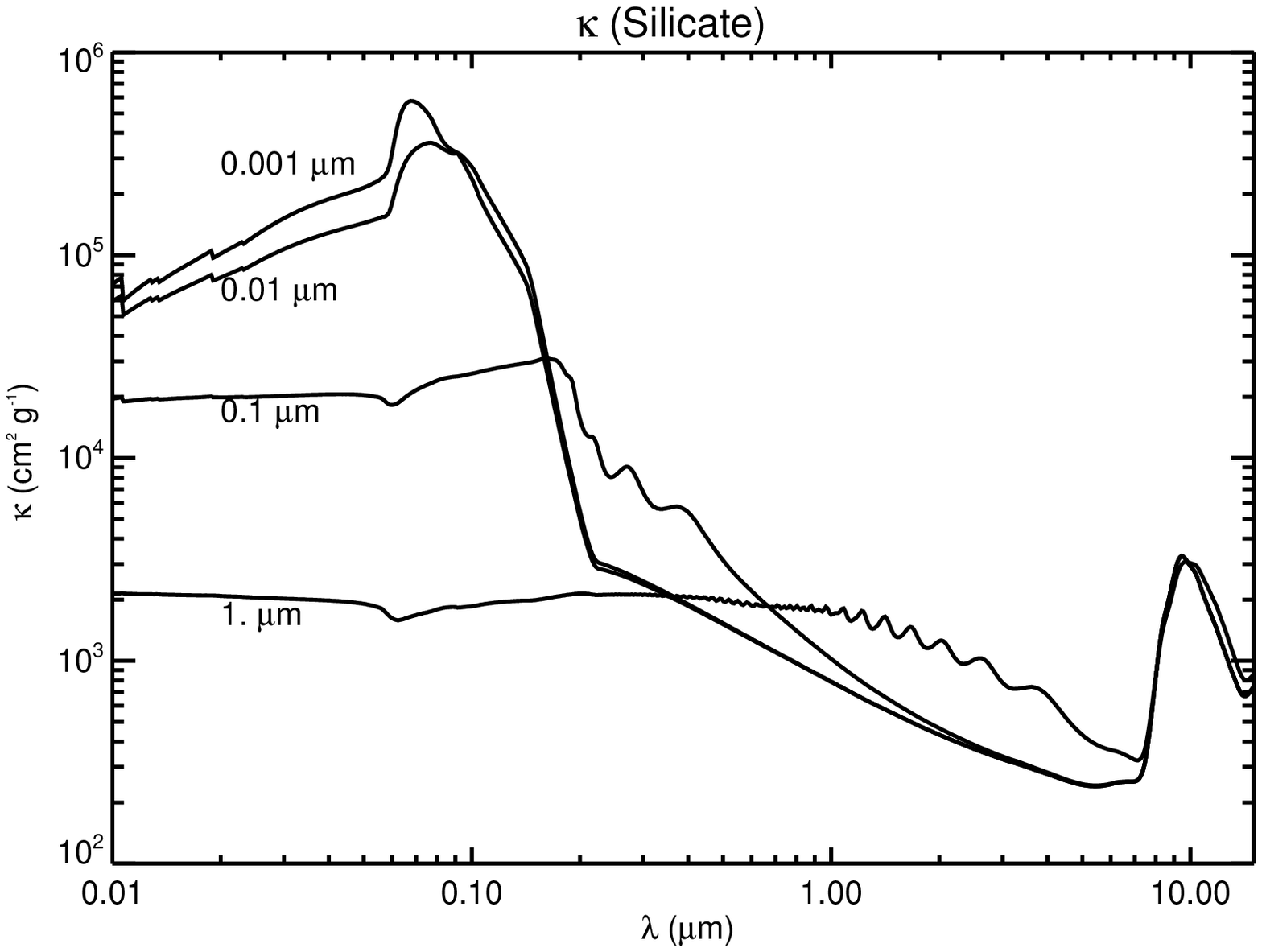}}
\subfigure{
\label{f4b}
\epsscale{0.5}
\plotone{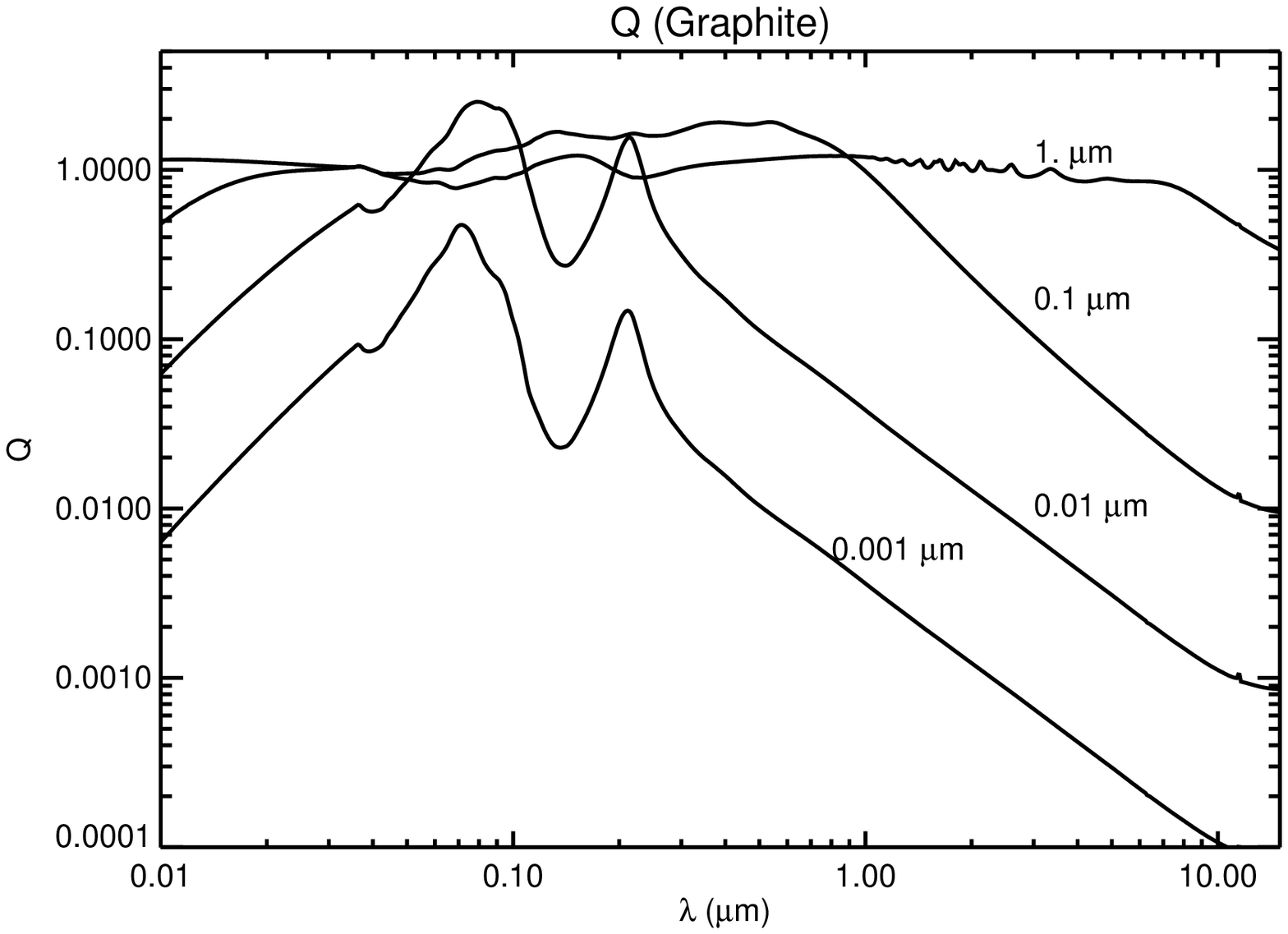}\plotone{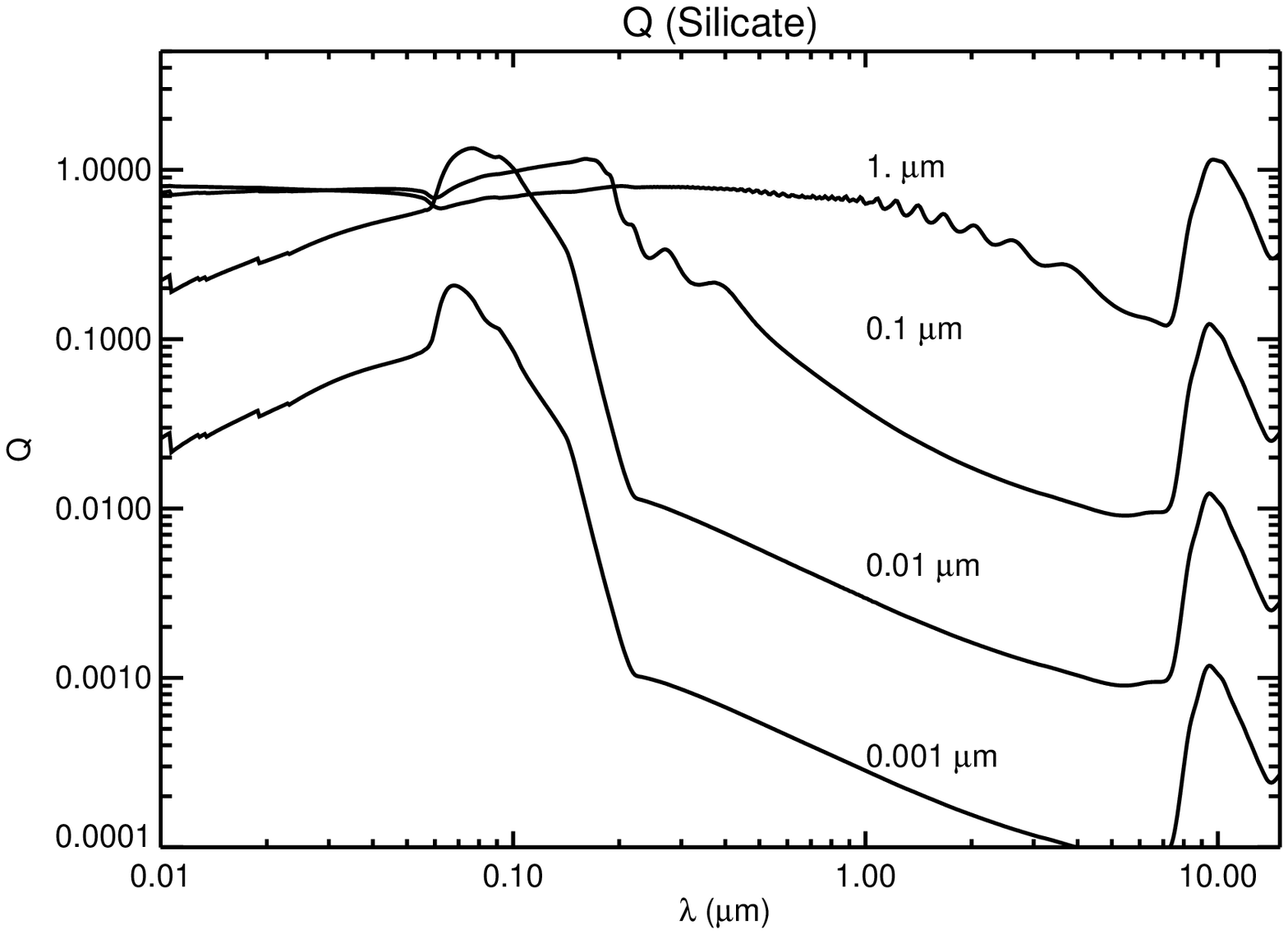}}
\caption{The dust opacity coefficient, $\kappa$, and emission efficiency, $Q$, for both graphite and silicate for several grain sizes as a function of wavelength.  The emission efficiency is given by the equation \ref{eqn:kappa}.  The dust opacity coefficient is nearly independent of grain radius assuming grains with radii less than the emitting wavelengths (i.e. $a \le 1$~\micron~at infrared wavelengths).  For large grains ($a > 1$~\micron), the dust can be approximated by a blackbody ($Q = 1$).}
\label{f4}
\end{center}
\end{figure*}

\begin{figure}
\begin{center}
\epsscale{1.2}
\plotone{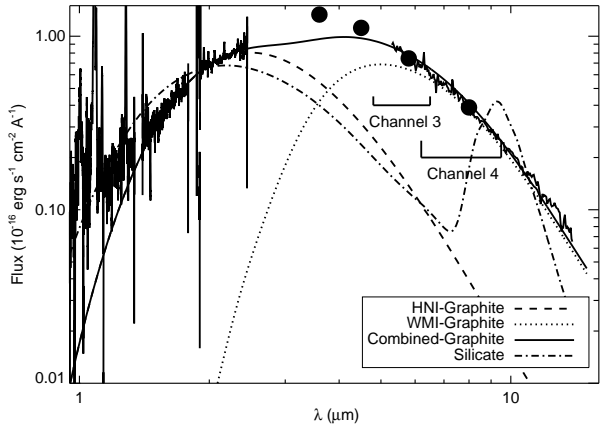}
\caption{
Best fits of the graphite and silicate models given by equation \ref{eqn:flux2} to the combined near- and mid-infrared data.  The lack of an emission feature at $\sim$9 \micron~immediately rules out the silicate model.  The combined spectrum is best fit by a multi-component model, consisting of both a ``hot'' ($\sim$800 K), near-infrared (HNI) and ``warm'' ($\sim$400 K), mid-infrared (WMI) component.
}
\label{f5}
\end{center}
\end{figure}

Given simple dust populations of a single size composed entirely of either silicate or graphite, Figure \ref{f4} plots the dust absorption coefficient and emission efficiency for several grain sizes of each composition, which is derived from Mie theory.  For the observed spectrum in Figure \ref{f3}, Figure \ref{f5} shows the best fit of equation \ref{eqn:flux2} with graphite and silicate models using IDL's {\tt MPFIT} function.  The lack of an emission feature at $\sim$9 \micron~immediately rules out any silicate grain contribution.  We therefore use only graphite models throughout the rest of this paper.  

We fit a multi-component model to the combined spectrum, where each component assumes a single dust mass, $M_{d}$, at a single temperature, $T_d$, composed of graphite of a single grain size.  The dust mass, temperature, and grain size of each component are all free parameters.  Figure \ref{f5} shows the optimal fit, which consists of two dominant components: a ``hot'' ($\sim$800 K), near-infrared (HNI) and ``warm'' ($\sim$400 K), mid-infrared (WMI) component.

Adding additional components to the fit tends to yield relatively small dust masses that do not improve $\chi^2$~by more than a couple percent.  This result has two implications.  First, two unique dust components exist, as opposed to a single dust component with a continuous temperature distribution.  Second, a hotter ($>1000$~K) third component, if it exists, does not significantly contaminate the near-infrared fits at these late times.  (At early times, a hot photospheric component is expected and is, in fact, observed to dominate the early $J$-band data in \citet{fox09}.)  We therefore only consider two-component fits throughout the rest of this article.

Table \ref{tab2} lists the dust masses, temperatures, and luminosities associated with various grain size combinations, but overall, grain size has little consequence on either these parameter values or the resulting goodness-of-fit ($\chi^2$).  The only size dependent variable in equation \ref{eqn:flux2} is $\kappa$, but Figure \ref{f4} shows the dust opacity coefficient for graphite is independent of grain radius at infrared wavelengths ($>$1 \micron) for grain sizes $<$1 \micron, which is typical for most grains.  Other methods (discussed in Sections \ref{sec3:shock} and \ref{sec3:echo}), however, can constrain the grain size.

Figure \ref{f6} plots the luminosities of both the HNI and WMI components on day 936 for 0.1 \micron~grains, along with the time-series evolution of SN 2005ip at near-infrared \citep{fox09} and visible luminosities \citep{smith09ip}.  \citet{smith09ip} show that as the photospheric component drops off over the first $\sim$100 days, an optical luminosity plateau,  $L_{\rm plateau}$, appears and continues throughout the extent of the observations.  This plateau arises from radiation generated by continuous shock interaction with the dense circumstellar medium, as opposed to an optical light echo powered by the peak supernova luminosity.  

Lacking mid-infrared observations, the data obtained prior to day 936 cannot distinguish between the multiple dust components.  The HNI component in Figure \ref{f5}, however, dominates the near-infrared observations (i.e., little contribution from the WMI component at near-infrared wavelengths), which plateau throughout the extent of the observations (see Figure \ref{f6}).  In fact, additional epochs of TripleSpec spectra obtained on days 862 and 895 show little evidence for spectral evolution at later times.  Given these observations, we assume throughout this paper that all near-infrared results predominantly represent the HNI component evolution.  For both the HNI and WMI components, the thermal emission arises from warm dust.  Disentangling the composition, origin, and heating mechanism of these dust components, however, requires a detailed analysis of potential heating mechanisms.

\section{Analysis: Dust Origin and Heating Mechanism}
\label{sec3:source}

\begin{deluxetable*}{ l c c c c c c c}
\tablewidth{0pt}
\tablecaption{Fitting Parameters \label{tab2}}
\tablecolumns{8}
\tablehead{
\colhead{Fit} & \colhead{$T_{\rm warm}$} & \colhead{$T_{\rm hot}$} & \colhead{$M_{\rm warm}$} & \colhead{$M_{\rm hot}$} & \colhead{$L_{\rm warm}$} & \colhead{$L_{\rm hot}$} & \colhead{$\chi^2$}\\
\colhead{} & \colhead{(K)} & \colhead{(K)} & \colhead{(\msolar)} & \colhead{(\msolar)} & \colhead{(\lsolar)} & \colhead{(\lsolar)} & \colhead{}
}
\startdata
1.0 \micron (warm) + 0.3 \micron (hot) & 568 & 1092 & $5.3 \times 10^{-3}$ & $4.5 \times 10^{-5}$ & $1.2 \times 10^{8}$ & $4.2 \times 10^{7}$ & 1.16\\
0.3 \micron (warm) + 0.3 \micron (hot) & 422 & 1061 & $2.6 \times 10^{-2}$ & $5.7 \times 10^{-5}$ & $1.1 \times 10^{8}$ & $4.6 \times 10^{7}$ & 1.13\\
0.1 \micron (warm) + 0.3 \micron (hot) & 467 & 1106 & $4.1 \times 10^{-2}$ & $4.1 \times 10^{-5}$ & $1.3 \times 10^{8}$ & $4.1 \times 10^{7}$ & 1.10\\
0.01 \micron (warm) + 0.3 \micron (hot) & 487 & 1113 & $4.2 \times 10^{-2}$ & $3.9 \times 10^{-5}$ & $1.3 \times 10^{8}$ & $4.0 \times 10^{7}$ & 1.10\\
0.001 \micron (warm) + 0.3 \micron (hot) & 487 & 1113 & $4.2 \times 10^{-2}$ & $3.9 \times 10^{-5}$ & $1.3 \times 10^{8}$ & $4.0 \times 10^{7}$ & 1.10\\
1.0 \micron (warm) + 0.1 \micron (hot) & 540 & 847 & $5.9 \times 10^{-3}$ & $5.9 \times 10^{-4}$ & $1.0 \times 10^{8}$ & $5.8 \times 10^{7}$ & 1.17\\
0.5 \micron (warm) + 0.1 \micron (hot) & 445 & 836 & $1.2 \times 10^{-2}$ & $6.8 \times 10^{-4}$ & $1.0 \times 10^{8}$ & $6.2 \times 10^{7}$ & 1.13\\
0.3 \micron (warm) + 0.1 \micron (hot) & 408 & 838 & $2.8 \times 10^{-2}$ & $6.7 \times 10^{-4}$ & $1.0 \times 10^{8}$ & $6.2 \times 10^{7}$ & 1.12\\
0.1 \micron (warm) + 0.1 \micron (hot) & 453 & 859 & $4.3 \times 10^{-2}$ & $5.2 \times 10^{-4}$ & $1.1 \times 10^{8}$ & $5.5 \times 10^{7}$ & 1.10\\
0.01 \micron (warm) + 0.1 \micron (hot) & 472 & 862 & $4.5 \times 10^{-2}$ & $5.0 \times 10^{-4}$ & $1.1 \times 10^{8}$ & $5.5 \times 10^{7}$ & 1.10\\
0.001 \micron (warm) + 0.1 \micron (hot) & 472 & 862 & $4.5 \times 10^{-2}$ & $5.0 \times 10^{-4}$ & $1.1 \times 10^{8}$ & $5.5 \times 10^{7}$ & 1.10\\
1.0 \micron (warm) + 0.01 \micron (hot) & 540 & 897 & $5.9 \times 10^{-3}$ & $6.4 \times 10^{-4}$ & $1.0 \times 10^{8}$ & $5.8 \times 10^{7}$ & 1.16\\
0.3 \micron (warm) + 0.01 \micron (hot) & 408 & 887 & $2.8 \times 10^{-2}$ & $7.3 \times 10^{-4}$ & $1.0 \times 10^{8}$ & $6.2 \times 10^{7}$ & 1.12\\
0.1 \micron (warm) + 0.01 \micron (hot) & 453 & 910 & $4.3 \times 10^{-2}$ & $5.7 \times 10^{-4}$ & $1.1 \times 10^{8}$ & $5.6 \times 10^{7}$ & 1.10\\
0.01 \micron (warm) + 0.01 \micron (hot) & 472 & 913 & $4.5 \times 10^{-2}$ & $5.5 \times 10^{-4}$ & $1.1 \times 10^{8}$ & $5.5 \times 10^{7}$ & 1.10\\
0.001 \micron (warm) + 0.01 \micron (hot) & 472 & 913 & $4.5 \times 10^{-2}$ & $5.5 \times 10^{-4}$ & $1.1 \times 10^{8}$ & $5.5 \times 10^{7}$ & 1.10
\enddata
\end{deluxetable*}

\begin{figure}[t]
\begin{center}
\epsscale{1.2}
\plotone{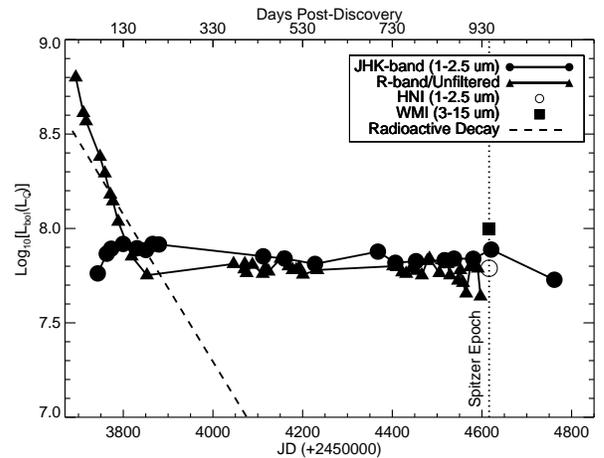}
\caption{
The evolution of SN 2005ip since discovery.  Included in this plot are near-infrared photometry from \citet{fox09}, unfiltered ($R$-band) photometry from \citet{smith09ip}, and the WMI and HNI luminosities on day 936 post-discovery derived from the best fits in Section \ref{sec3:dust} and shown in Figure \ref{f5}.  The radioactive decay for a typical (0.1 \msolar~of $^{56}$Ni) supernova is overplotted (dashed line).  
}
\label{f6}
\end{center}
\end{figure}

Section \ref{sec3:intro} summarized likely origins and heating mechanisms for late-time infrared dust emission, distinguishing scenarios involving newly formed versus pre-existing dust.  \citet{smith09ip} provide spectroscopic evidence for new dust condensation in the ejecta from days 60-170 and in the cool, dense shell of post-shocked gas at days $>$413.  This dust formation timeline closely corresponds to the evolution of the near-infrared light curve in Figure \ref{f6}.  The relative contribution of this newly formed dust to the HNI and WMI components, however, is not immediately clear, but can be addressed by the $Spitzer$~mid-infrared photometry and spectroscopy presented here in the context of both pre-existing dust scenarios: shock heating and an infrared echo.

\subsection{The Blackbody and Shock Radii}
\label{sec3:radii}

For both shock heating and an infrared echo, the blackbody and shock radii respectively serve as useful reference points.  The blackbody radius, given as $r_{bb} = \big(\frac{L_{bb}}{4 \pi \sigma T_{bb}^4}\big)^\frac{1}{2}$~where $\sigma$~is Stefan-Boltzmann's constant, defines the minimum shell size of an observed dust component.  In the case of SN 2005ip, blackbody fits ($Q=1$) of the combined spectrum on day 936 in Figure \ref{f5} yields blackbody radii of $r_{bb} {\rm (WMI)} \approx $\radbbwarm~and $r_{bb} {\rm (HNI)} \approx $\radbbhot.  

The shock radius, given as $r_s = v_s t$ for a constant velocity, $v_s$, defines the maximum radius that the forward shock can travel in a time, $t$.  The shock velocity can be derived from the optical emission line widths.  For SN 2005ip, \citet{smith09ip} observe broad emission lines corresponding to radial velocities of $\sim$15,000 \kms~(0.05$c$) through $\sim$900 days post-discovery, yielding a maximum shock radius on day 936 of $r_{s1} \approx 1.25 \times 10^{17}$~cm (0.125 ly).  At the same time, the intermediate width lines correspond to slower shock velocities of $\sim$1000 \kms~(0.003$c$) through $\sim$900 days post-discovery.  \citet{chugai94} propose that two unique shock velocities can coexist if the progenitor's wind is clumpy or asymmetric, as opposed to homogeneous and spherical.  A relatively rarefied wind allows uninhibited shocks to maintain the observed high velocities, while a much slower shock propagates through the denser regions.

\begin{figure*}
\begin{center}
\epsscale{0.5}
\subfigure{\label{f7a} \plotone{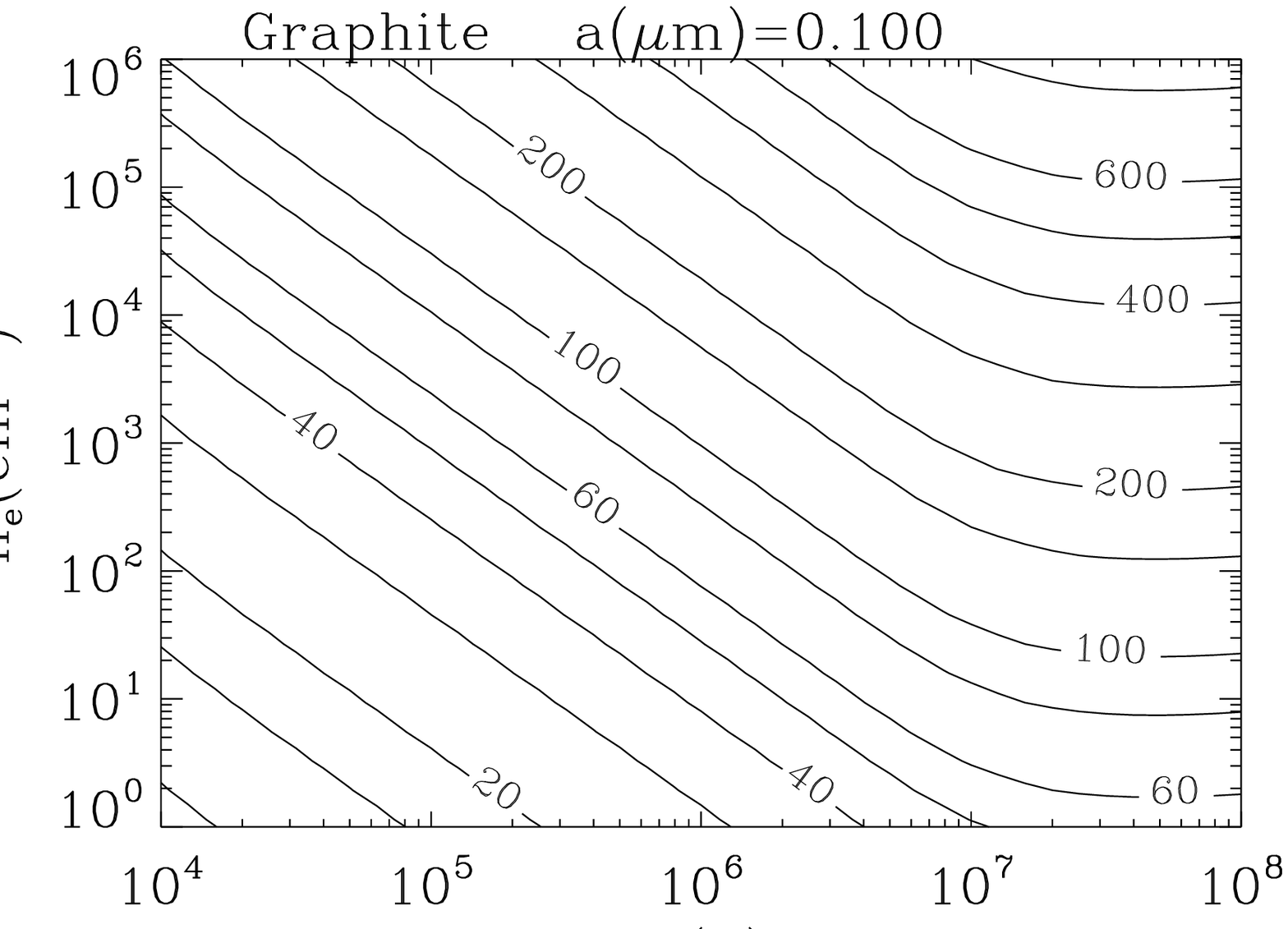}}
\subfigure{\label{f7b} \plotone{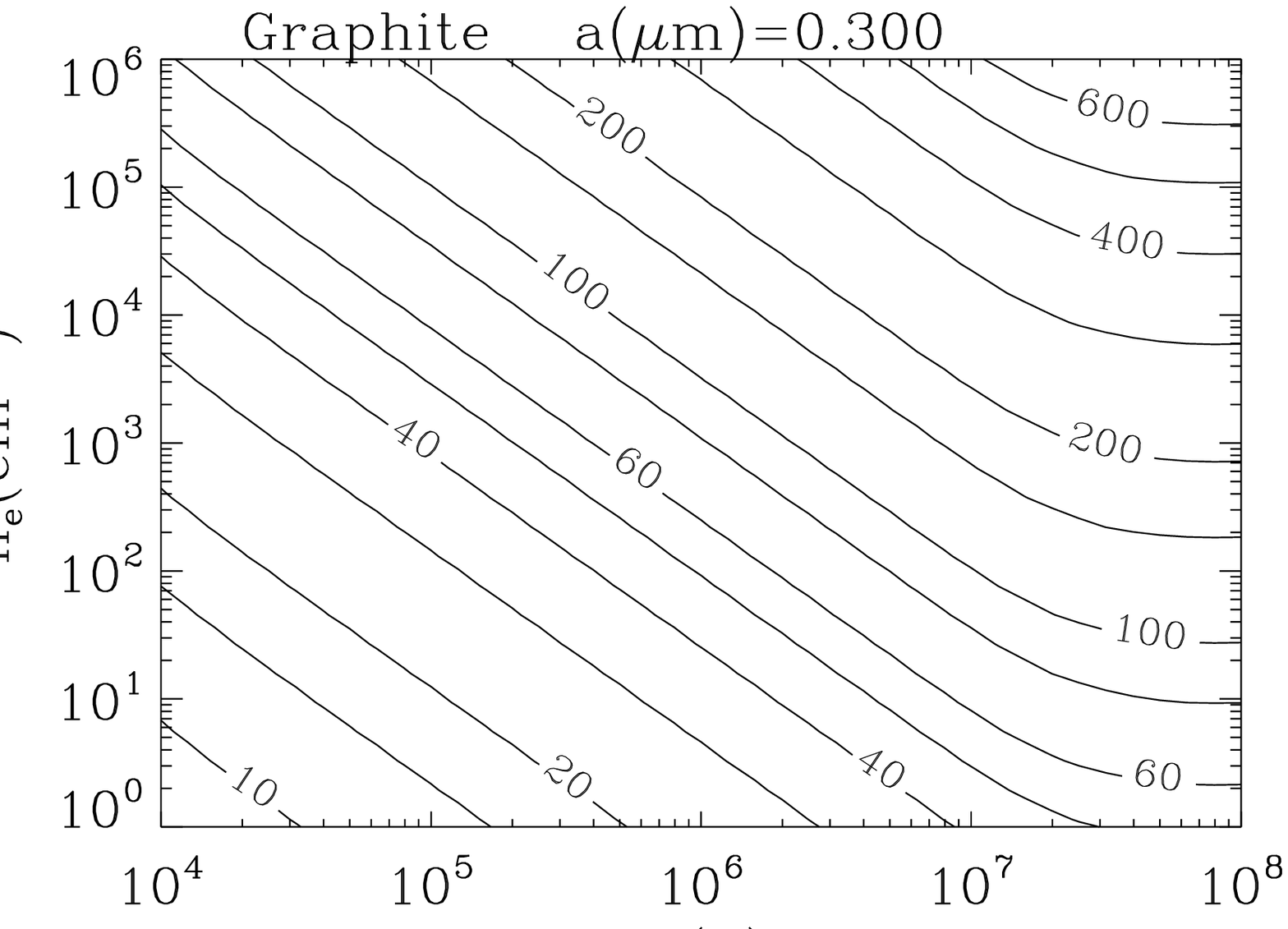}}
\caption{Post-shock equilibrium dust temperatures ($T_d$~(K)) as a function of post-shock electron density, $n_{\rm e}$, and temperature, $T_{\rm e}$.  For typical post-shock electron temperatures, $T_e \approx 10^7$~K, the grains become increasingly transparent to the incident electrons, and the dust temperature is only a function of the electron density.  
}
\label{f7}
\end{center}
\end{figure*}

Since a shock will typically destroy any pre-existing dust, understanding the shock propagation is essential for modeling the evolution of any late-time infrared emission.  If, for example, the denser regions are concentrated in a homogeneous equatorial disk, only the slower shocks need be considered as they propagate contiguously throughout the disk.  Alternatively, if the denser regions are concentrated in clumps, the situation is more complicated.  The fastest shocks will reach the clumps first, at which point the shock velocity drops.  The average shock velocity is a function of the clump filling factor and distribution.  Since the distribution of the dense regions is not well known, we consider a second shock radius for which we adopt a more modest shock velocity of $v_s \sim$5000 \kms, which is consistent with the intermediate component of many Type IIn supernovae, \citep[e.g.][]{salamanca02, smith08tf, smith09gy, green08}.  This velocity yields an upper limit on a second shock radius on day 936 is $r_{s2} \approx 4 \times 10^{16}$~cm (0.04 ly).

Both the HNI and WMI shells are optically thin, thereby confirming the assumptions presented in Section \ref{sec3:dust}.  The optical depth of each shell can be written as,
\begin{equation}
\tau = \frac{M_d}{4 \pi r^2}  \kappa_{\rm avg},
\label{eqn:tau}
\end{equation}
where $\kappa_{\rm avg} = 435$~cm$^2$~g$^{-1}$~is the absorption coefficient for graphite averaged over $1 - 15$~\micron.  For 0.1 \micron~grains, Table \ref{tab2}~shows $M_d (HNI) \approx 5.2 \times 10^{-4}$~\msolar~and $M_d (WMI) \approx 4.3 \times 10^{-2}$~\msolar.  Given the minimum radii, $r > r_{bb}(HNI) \approx $\radbbhot~and $r > r_{bb}(WMI) \approx $\radbbwarm, equation \ref{eqn:tau} yields $\tau < $ 0.6 and 1.3, respectively.


\subsection{Shock Heating}
\label{sec3:shock}

In the shock heating scenario, hot electrons in the post-shock environment collisionally heat pre-existing dust grains.  \citet{dwek87} and \citet{dwek08} provide a detailed description of this process for silicate dust grains and present post-shock equilibrium dust temperatures as a function of post-shock electron density, $n_{\rm e}$, and temperature, $T_{\rm e}$.  Figure \ref{f7} presents a similar analysis for graphite grains. 

For shock heating to occur, the forward shock must have sufficient time to reach the pre-existing dust grains.  As discussed in further detail in Section \ref{sec3:echo}, the peak supernova luminosity of SN 2005ip ($L_{peak} \approx 10^{9}$~\lsolar) vaporizes all dust grains out to a radius, $r_{\rm evap} \approx 10^{16}$~cm (0.01 ly).  For the fastest observed shock velocities ($v_s\sim$15,000 \kms), a shock would require $\sim$70 days to reach the evaporation radius, which is consistent with the earliest observations of the HNI component (see Figure \ref{f6}).  The earliest observation of the WMI component does not occur until the $Spitzer$~observations on day 936, by which point even the slower shocks would have crossed the evaporation radius.  

An independent calculation of the shocked dust mass provides a useful consistency check.  Assuming the hot, post-shocked gas heats the dust shell, estimates of the total gas mass will determine the dust-to-gas mass ratio.  The upper limit on the volume of the emitting shell is given at 
\begin{equation}
V_{\rm shell}  = 4 \pi r_{\rm s}^2 \Delta r_{\rm s},
\label{eqn:volume}
\end{equation}
where $r_{\rm s}$ is the radius of the shock with velocity, $v_{\rm s}$, at an age, $t$,
and $\Delta r_{\rm s}$ is the shell thickness defined by the distance traveled by the shock over the grain sputtering lifetime, $\tau_{\rm sputt}$, 
\begin{equation}
\Delta r_{\rm s} = \frac{1}{4}v_{\rm s} \tau_{\rm sputt},
\end{equation}
provided that $\tau_{\rm sputt} < (t, \tau_{\rm cool})$, where $\tau_{\rm cool}$~is the radiative cooling time-scale.  The factor of $\frac{1}{4}$ comes from the shock jump conditions.  For typical post-shock gas temperatures ($> 10^6$~K), \citet{dwek92} give the sputtering lifetime for a grain size, $a$, and gas density, $n_g$, as
\begin{equation}
\tau_{\rm sputt} ({\rm yr}) \approx 10^6 \frac{a(\micron)}{n_g({\rm cm^{-3}})}.
\label{eqn:sputter}
\end{equation}
The total gas mass of the emitting dust shell is therefore
\begin{equation}
\label{eqn:mg}
M_{\rm g} = n_g m_{\rm H} V_{\rm shell}.
\end{equation}
Combining equations \ref{eqn:volume} - \ref{eqn:mg} yields
\begin{equation}
\label{eqn:dustmass}
M_{\rm g} ({\rm M_{\odot}}) \approx 8.3 \times 10^{-5} \bigg(\frac{v_{\rm s}}{{\rm 1000~\kms}}\bigg)^3 \bigg(\frac{t}{{\rm yr}}\bigg)^2 \bigg(\frac{a}{{\rm \micron}}\bigg),
\end{equation}
which reveals the mass of the shocked gas is independent of the grain density. 

Assuming a dust-to-gas mass ratio expected in the H-rich envelope of a massive star, $Z_d = \frac{M_d}{M_g} \approx 0.01$, gives the expected dust mass.  For the maximum observed shock velocity, $v_s < $ 15,000~\kms, and age, $t = 936$~days, Table \ref{tab3} compares the predicted dust mass to the observed mass listed in Table \ref{tab2}.  Assuming shock heating is also responsible for the HNI component at early times, Table \ref{tab4} compares the predicted HNI dust mass at an age $t = 70$~days to the observed mass listed in Table \ref{tab2}.  Although no measurement of the dust mass exists on day $\sim$70, the relatively constant HNI luminosity (see Figure \ref{f6}) and temperature (see \citet{fox09}) suggest a relatively constant HNI mass throughout the extent of the observations.  

For both the WMI component on day 936 and the HNI component on day 70, only large grains ($a>0.3$~\micron) can reproduce the observed dust masses.  These grain radii are large compared with typical grain sizes observed in supernova shocks \citep{dwek08}.  Furthermore, upper limits were assumed for both the shock velocity (see Section \ref{sec3:radii}) and dust-to-gas mass ratio \citep{williams06}.  Lower values would require even larger grain sizes to reproduce the observed dust masses.  These results likely rule out shock heating.

\subsection{Possible Emission From an IR Echo}
\label{sec3:echo}

For an infrared echo scenario, the supernova luminosity heats a shell of dust at a radius, $r$, to a peak temperature, $T_d$.  This outer shell may be pre-existing at the time of the supernova explosion or it may form when the peak supernova luminosity creates a vaporization cavity.  In either case, light travel time effects cause the thermal radiation from the dust grains to reach the observer over an extended period of time, thereby forming an `IR echo' \citep{bode80,dwek83}.  The infrared luminosity plateau occurs on year long time scales, corresponding to the light travel time across the inner edge of the dust shell.  As dust cools from the peak temperature, it will contribute flux at longer wavelengths.  As noted in Section \ref{sec3:dust}, however, the SN 2005ip spectrum is best fit by two components, as opposed to a continuous temperature distribution.  Therefore, this analysis assumes a simple light echo model that is dominated by flux from only the warmest dust with a single temperature, $T_d$.

\begin{deluxetable}{ c c c c c }
\tablewidth{0pt}
\tablecaption{Shock Heating Mass Predictions at $t=936$~Days for $v_s$~=~15,000 \kms \label{tab3}}
\tablecolumns{5}
\tablehead{
\multirow{3}{*}{a (\micron)}
 & \multicolumn{2}{c}{WMI} & \multicolumn{2}{c}{HNI}\\
 & \colhead{M$_d$~(\msolar)} & \colhead{M$_d$~(\msolar)} & \colhead{M$_d$~(\msolar)} & \colhead{M$_d$~(\msolar)}\\
 & \colhead{Predicted} & \colhead{Observed} & \colhead{Predicted} & \colhead{Observed}
}
\startdata
0.01 & 1.8e-4 & 4.5e-2 & 1.8e-4 & 5.7e-4 \\
0.1 & 1.8e-3 & 4.3e-2 & 1.8e-3 & 5.2e-4 \\
0.3 & 5.5e-3 & 2.8e-2 & 5.5e-3 & 4.1e-5 \\
0.5 & 9.2e-3 & 1.2e-2 & - & - \\
1.0 & 1.8e-2 & 5.9e-3 & - & -
\enddata
\end{deluxetable}

The equilibrium dust temperature is set by balancing the energy absorbed and emitted by the dust grains,
\begin{equation}
\label{eqn:balance}
L_{\rm abs} = L_{\rm rad},
\end{equation}
where, for a single dust grain,
\begin{eqnarray}
\label{eqn:labs}
L_{\rm abs}  & = & 4 \pi r_{\rm SN}^2 \frac{\pi a^2}{4 \pi r^2} \int{\pi B_\nu(T_{\rm SN}) Q_{\rm abs}(\nu) d\nu} \nonumber \\
	      & = & \frac{L_{\rm bol}}{\sigma T_{\rm SN}^4} \frac{\pi a^2}{4 r^2} \int{B_\nu(T_{\rm SN}) Q_{\rm abs}(\nu) d\nu}
\end{eqnarray}
and 
\begin{eqnarray}
\label{eqn:lrad}
L_{\rm rad} & = & 4 \pi a^2 \int{\pi B_\nu (T_d) Q_{\rm abs}(\nu) d\nu} \nonumber\\
 	            & = & \frac{16}{3} \pi \rho a^3 \int{B_\nu (T_d) \kappa(\nu) d\nu}
\end{eqnarray}
where $r_{\rm SN}$~is the effective supernova emitting radius, $L_{\rm bol}$~is the UV-optical luminosity, and $T_{\rm SN}$ is the effective supernova blackbody temperature.  $L_{\rm bol}$ follows from Equations \ref{eqn:balance}, \ref{eqn:labs}, and \ref{eqn:lrad},
\begin{equation}
\label{eqn:lbol}
L_{\rm bol}  = \frac{64}{3} \rho a r^2 \sigma T_{\rm SN}^4 \frac{\int{B_\nu (T_d) \kappa(\nu) d\nu}}{\int{B_\nu(T_{\rm SN}) Q_{\rm abs}(\nu) d\nu}}.
\end{equation}
$L_{\rm bol}$~depends on the grain radius because although the dust opacity coefficient, $\kappa$, in Figure \ref{f4} is independent of grain radius for thermal emission at longer wavelengths, it does depend on grain radius for absorption at shorter wavelengths (e.g., UV and optical).

\begin{figure*}
\begin{center}
\epsscale{0.6}
\subfigure[]{\label{f8a} \plotone{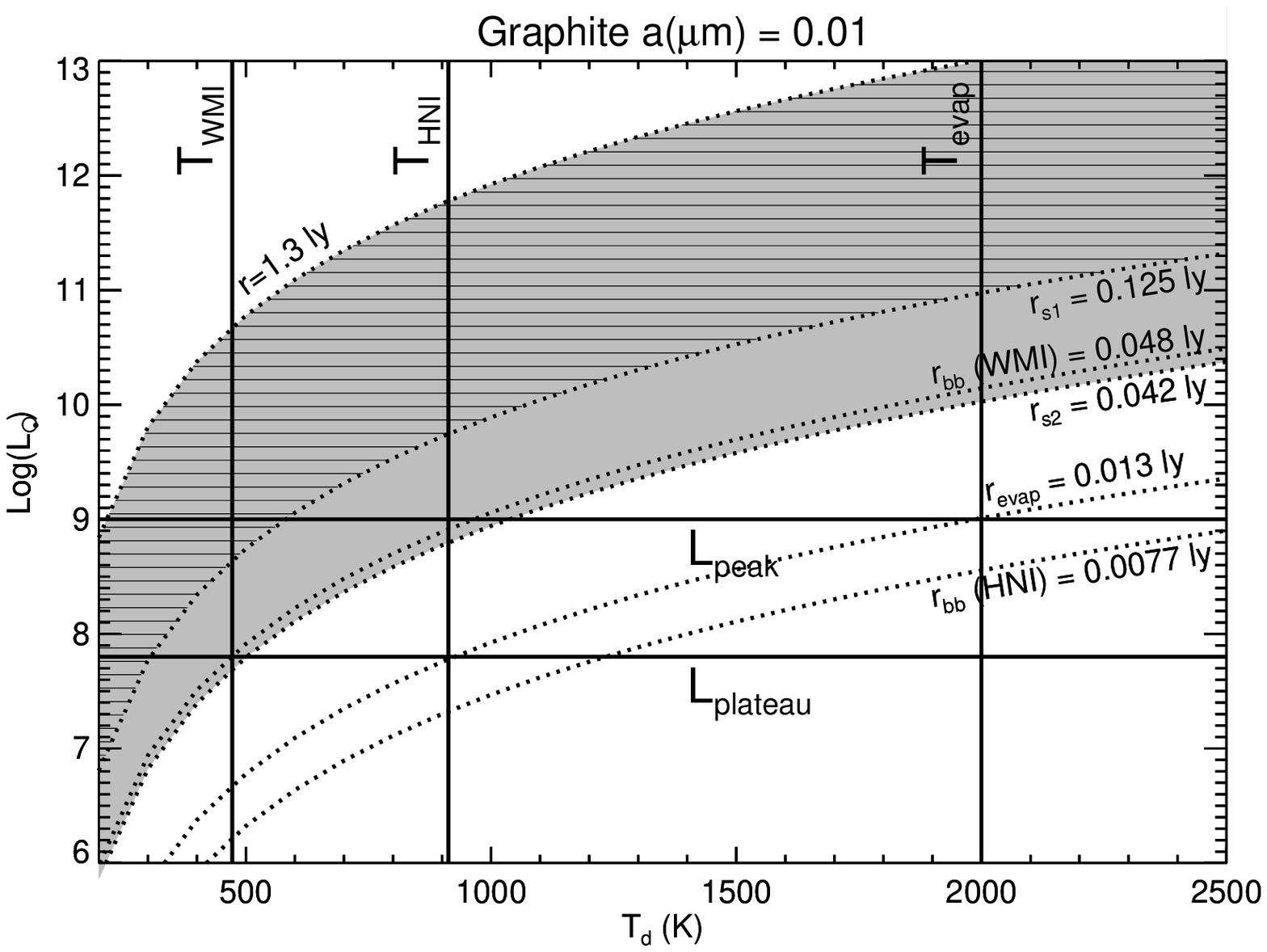}}\\
\subfigure[]{\label{f8b} \plotone{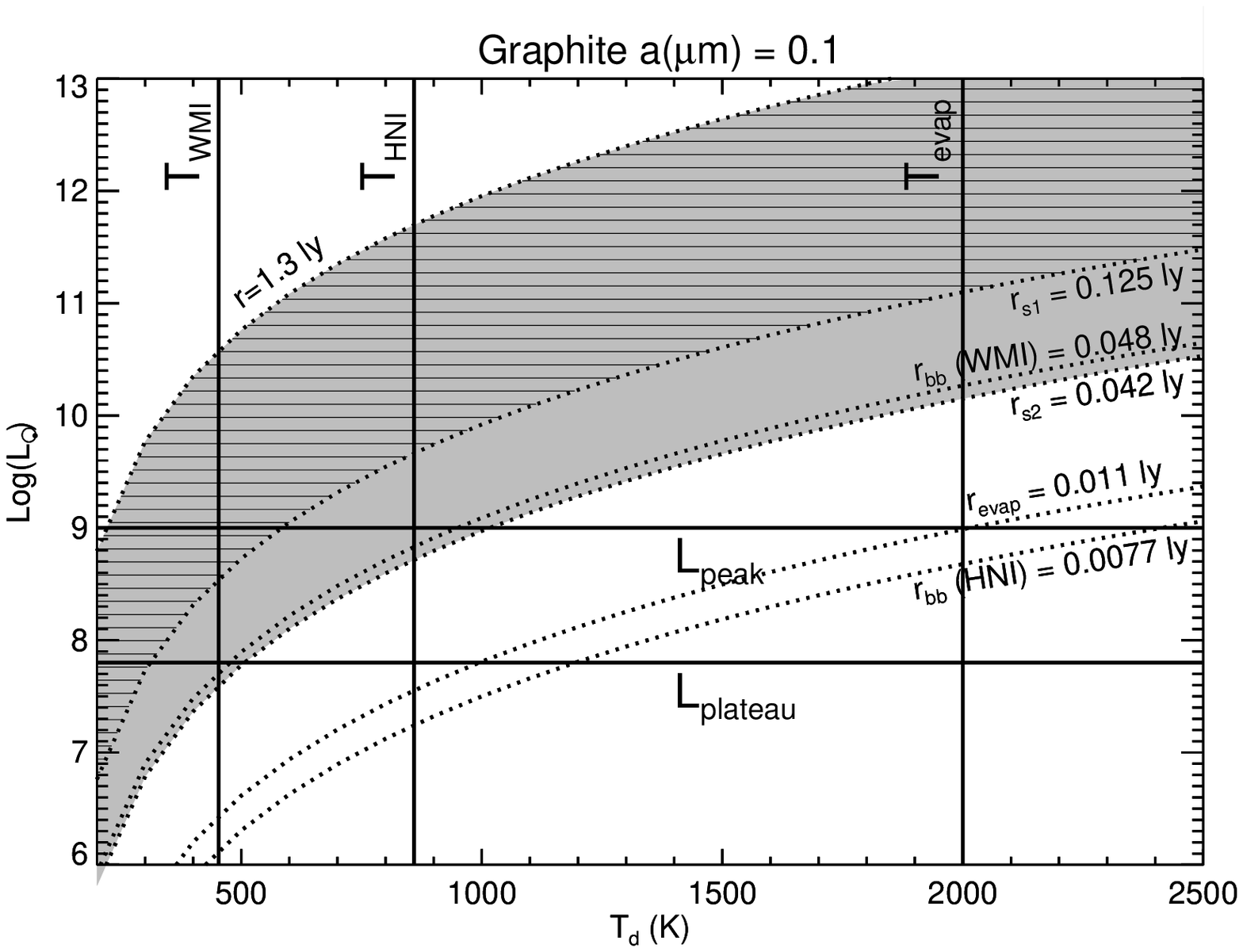}}
\subfigure[]{\label{f8c} \plotone{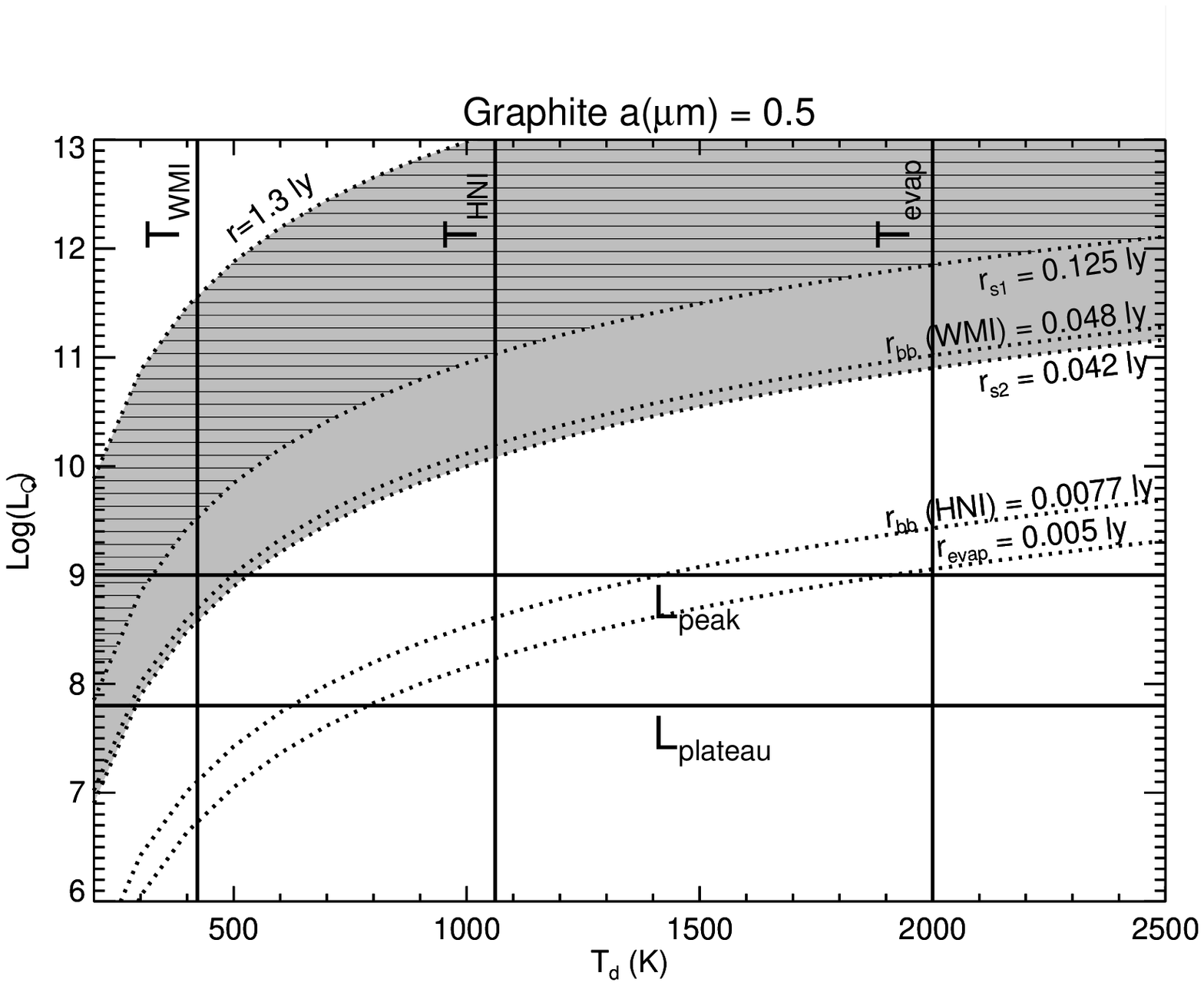}}
\caption{
Contours of the dust shell size, $r$, plotted as a function of both the supernova luminosity, $L_{\rm bol}$, and dust temperature, $T_d$, given by equation \ref{eqn:lbol}.  Several grain sizes are considered for dust with a graphite composition.  The vertical lines show the observed graphite dust temperatures listed in Table \ref{tab2} and the approximate vaporization temperature of graphite dust, $T_{\rm evap} \approx 2000$~K.  The horizontal lines show both the observed peak, $L_{\rm peak}$, and late-time optical plateau, $L_{\rm plateau}$, luminosities from Figure \ref{f6}.  The shaded regions highlight the dust shell radii allowed by the constraints.  The shock radius sets the lower limit as any dust within this radius would be independently heated or destroyed by the forward shock (see Section \ref{sec3:shock}).  Both shock radii described in Section \ref{sec3:radii} are considered.  The horizontal line fill distinguishes $r_{s1}$~from $r_{s2}$.  The near-infrared plateau time-scale sets the upper limit.  Figure \ref{f6} shows the plateau extends for at least $\sim$2.6 years and still shows little sign of declining.  The minimum upper limit is therefore 1.3 light years, although it may certainly be larger.  Section \ref{sec3:echo} considers three possible echo scenarios.
}
\label{f8}
\end{center}
\end{figure*}

Using equation \ref{eqn:lbol}, Figure \ref{f8} plots contours of the shell size, $r$, as a function of both luminosity, $L_{\rm bol}$, and observed dust temperature, $T_d$.  The luminosity is treated as a central point source, assuming the emitting region is internal to a spherically symmetric dust shell.  Several grain sizes are considered for dust with a graphite composition.  Although the calculation assumes $T_{\rm SN} \approx$ 10,000 K, the result is fairly insensitive to this choice.  The vertical lines show the observed graphite dust temperatures listed in Table \ref{tab2} and the approximate vaporization temperature of graphite dust, $T_{\rm evap} \approx 2000$~K.  The horizontal lines show both the observed peak, $L_{\rm peak}$, and late-time optical/infrared plateau, $L_{\rm plateau}$, luminosities from Figure \ref{f6}.  

\begin{deluxetable}{ c c c c c }
\tablewidth{0pt}
\tablecaption{Shock Heating Mass Predictions at $t=70$~Days for $v_s$~=~15,000 \kms \label{tab4}}
\tablecolumns{3}
\tablehead{
\multirow{3}{*}{a (\micron)}
 & \multicolumn{2}{c}{HNI}\\
 & \colhead{M$_d$~(\msolar)} & \colhead{M$_d$~(\msolar)}\\
 & \colhead{Predicted} & \colhead{Observed}
}
\startdata
0.01 & 1.0e-6 & 5.7e-4 \\
0.1 & 1.0e-5 & 5.2e-4 \\
0.3 &  3.1e-5 & 4.1e-5 \\
\enddata
\end{deluxetable}

The shaded regions highlight the dust shell radii allowed by the constraints.  The shock radius sets the lower limit as any dust within this radius would be independently heated or destroyed by the forward shock (see Section \ref{sec3:shock} above).  (For the WMI component, the blackbody radius, $r_{bb}(WMI)$,~actually sets the minimum radius, as described in Section \ref{sec3:radii}.)  Both the fast, $r_{s1}$, and slow, $r_{s2}$, shocks described in Section \ref{sec3:radii} are considered, distinguished by the hashed region.  The near-infrared plateau time-scale sets the upper limit.  Figure \ref{f6} shows the plateau extends for at least $\sim$2.6 years and still shows little sign of declining.  The current upper limit is therefore set by a ``plateau radius,'' $r_p$, of 1.3 light years, although the final upper limit will only be determined once the light-curve plateau begins to decline.  Should the plateau extend even longer than 2.6 years the larger implied radius only makes a light echo less likely.

Three possible light echo scenarios exist:\\
{\bf 1)}  In the first scenario, a sphere of dust completely encompasses the progenitor at the time of the explosion.  The peak luminosity vaporizes all dust within a radius, $r_{\rm evap}$, and warms the inside of the remaining dust shell to nearly the vaporization temperature ($T_{\rm evap}~\approx~2000$~K).  Figure \ref{f8a} shows that the observed peak optical luminosity, $L_{\rm peak}$, yields a vaporization radius of $r_{\rm evap} \sim 0.013$~light years for $a=0.01$~\micron~graphite grains and an even smaller radius for larger grains (see Figures \ref{f8b} and \ref{f8c}).  These small evaporation radii, however, are inconsistent with the observations as a shell of this size cannot produce an infrared echo on 3 year time scales.  Furthermore, these evaporation radii are smaller than both shock radii.  Even if the actual peak luminosity were a factor of 5 larger than observed ($L_{\rm est} \sim5\times10^{9}$~\lsolar), the vaporization radius would be insufficient to account for the observations.  We therefore rule out this scenario for both the HNI and WMI components.

{\bf 2)}  The dust shell inner limit need not lie exactly at the vaporization radius.  If the progenitor underwent an eruption many years before the supernova, the dust shell may lie at larger radii.  In this second scenario, the peak luminosity, $L_{\rm peak}$, heats the dust shell inner radius to only the observed temperature.  The light echo duration therefore defines the minimum cavity radius (i.e., $r_p$=1.3 light years).  Figure \ref{f8} shows that a minimum peak luminosity of $L_{\rm bol} > 5 \times 10^{10}$~\lsolar~is required to heat a dust shell of 1.3 light years in radius to the observed WMI temperature, while a minimum peak luminosity of $L_{\rm bol} > 5 \times 10^{11}$~\lsolar~is required for the HNI component.  A larger shell radius requires even larger peak luminosities.  

These required peak luminosities are significantly larger than the observed peak luminosity.  The true peak luminosity of SN 2005ip, however, is not well constrained.  The earliest R-band photometry was obtained 14-days post-discovery, and the discovery occurred a few weeks following the actual explosion so that the peak luminosity was likely several times larger than the observed early-time $R$-band luminosity ($L_{\rm est} \sim5\times10^{9}$~\lsolar).  Still, while a significant amount of optical absorption might be expected by large amounts of pre-existing dust, significant reddening was not observed \citep{smith09ip} and a peak luminosity $> 10^{10}$~\lsolar~is unlikely from this extrapolation.  

The shock breakout in the minutes to hours following the supernova explosion may reach peak luminosities $> 10^{11}$~\lsolar~\citep{soderberg08,rest09}, but no such breakout was observed for SN 2005ip.  A peak luminosity $>10^{11}$~\lsolar~would have made SN 2005ip one of the most luminous core-collapse events ever observed \citep{quimby07,rest09}, but figure 11 of \citet{rest09} suggests SN 2005ip is nearly an order of magnitude fainter than SNe 2006gy 2008es, 2005ap, and 2003ma, especially at early times.  Furthermore, optical emission from the late-time circumstellar interaction successfully accounts for the observed dust temperatures (see below).  These reasons rule out a light echo driven by the peak supernova luminosity for both the HNI and WMI components.

{\bf 3)} A final scenario considers a pre-existing dust shell similar to scenario 2 above, but in this case the shell's inner radius is located at an intermediate radius between the shock and plateau radii (as defined on day 936).  The late-time optical emission, $L_{\rm plateau}$, continuously heats the dust shell to the observed temperature.  This scenario is not so much a traditional infrared echo as it is a reprocessing of the optical emission by the dust.  (Some authors refer to this as a circumstellar shock echo \citep{gerardy02}.)  If the circumstellar interaction occurs on a time scale greater than the light travel time across the dust shell, the shell radius does not set the infrared plateau length.  The observed flux therefore accounts for the entire shell.

Figures \ref{f8a}~and \ref{f8b}~show that for both $a=0.01$~and 0.1 \micron~graphite grains, $L_{\rm plateau}$ can heat a dust shell of radius $r \approx 0.01$~ly to $T_{\rm HNI}$~and a shell of radius $r \approx 0.05$~ly to $T_{\rm WMI}$.  For $a=0.5$~\micron~graphite grains, Figure \ref{f8c} shows that $L_{\rm plateau}$ can only heat a dust shell of radius $r < 0.005$~ly to $T_{\rm HNI}$~and a shell of radius $r < 0.04$~ly to $T_{\rm WMI}$.  In this scenario, both the fast ($r_{S1} = 0.125$~ly) and slow ($r_{\rm S2} = 0.042$~ly) shock radii are larger than the HNI shell radii by day 936, ruling out a light echo of this sort for the HNI component.  The same is true for the WMI shell composed of larger grains ($a > 0.5$~\micron).  In the case of the WMI shell composed of smaller grains ($a < 0.1$~\micron), however, the slower shock has not yet reached the shell radius ($r \approx 0.05$~ly).  Not only is this scenario possible for the WMI component, but the radius is consistent with the WMI blackbody radius ($r_{\rm bb}$ (WMI) = 0.048 ly).  Furthermore, the grain sizes are typical of those observed in other supernova circumstellar environments \citep{dwek08}.

Predictions for the light curve evolution can be made from this model.  Dust that remains at radii beyond the slower shock radius, $r_{S2}$, will continue to radiate and contribute to the light echo plateau.  Although the exact distribution of these dense, dusty regions is not well-known, this scenario suggests they must be distributed in such a way that the fastest shocks do not interact with the dust.  As described in Section \ref{sec3:radii}, these dense regions may be concentrated in either an equatorial disk or clumps.  The clumps must have a large filling factor if the fastest shocks are not to interact with a significant portion of the dust.  As the slower shocks continues to expand, however, they will ultimately destroy the dust.  Assuming the dust lies at a radius consistent with 0.1 \micron~grains (see parameters in Table \ref{tab2}), the WMI flux will begin to decrease at $t \approx \frac{r ({\rm WMI})}{v_s} \sim 2775$~days post-discovery and continue to decrease as a function of the emitting shock radius and the dust distribution.  While this scenario is entirely consistent with the WMI observations at the current time, mid-infrared observations at later epochs can reveal the accuracy of this model's predictions.

\section{Discussion}
\label{sec3:multi}

\subsection{The HNI Component}
The above HNI component analysis rules out the possibility of both pre-existing dust scenarios (i.e., shock heating or infrared echo).  Only condensation models (i.e., in the ejecta or cool, dense shell) remain as viable scenarios for explaining the origin of the HNI component.  \citet{smith09ip} confirm new dust formation via extinction in the broad ($\sim$15,000 \kms) H$\alpha$~wings between days $\sim$60-170 and in the intermediate ($\sim$2000 \kms) He I wings at days $>$413, but the location of this new dust remains ambiguous.  The extinction in the broad and intermediate components suggest new dust in the fast ejecta at early times and post-shock cool, dense shell at later times, respectively.  

To form dust at the high ejecta velocities, however, is difficult due to the lack of heavy metals traveling at these speeds.  If the dust formed in the cool, dense shell at early times, as suggested in the case of SN 2006jc \citep{mattila08}, this would explain the attenuation of the broad H$\alpha$~line observed by \citet{smith09ip} without having to invoke dust formation at high velocities, but would fail to explain the lack of observed intermediate width He I lines produced by the circumstellar interaction.  \citet{smith09ip} propose one alternative scenario in which the dust may form in the post-shock gas of individual clumps, which then become incorporated into the expanding fast ejecta when a clump is eventually destroyed.  This scenario, however, requires several assumptions, most of which rely on unknown clumping properties.  While the available observations limit the ability to isolate the precise region of dust formation, the relatively flat near-infrared flux in Figure \ref{f6} suggests a majority of the dust contributing to the HNI flux must have formed at early epochs.

Newly formed dust cannot reproduce the observed HNI luminosity plateau solely by cooling from condensation to the observed temperature $T_{\rm HNI} \approx 800$~K.  Given an average energy per particle $\epsilon = C_g \Delta T_d$ for the specific heat for graphite $C_g$ (given by \citet{draine01}) and $\Delta T \approx 1200$~K, an unsustainable mass condensation rate of $\dot{M} = L_{\rm HNI} / \epsilon \approx 100$ \msolar~day$^{-1}$ would be necessary to reproduce the observed flux (see a more detailed explanation of this argument for the case of SN 2006jc in \citet{fox09}).
 
Instead, an alternative heating mechanism must power the thermal emission from this newly formed dust.  Figure \ref{f6} shows that radioactive heating is insufficient to power the late-time near-infrared emission.  While heating by the reverse shock is possible, a similar analysis as performed in Section \ref{sec3:shock} suggests this scenario is unlikely because only large grains ($a > 0.5$~\micron) can reproduce the observed flux.  More likely, the optical luminosity generated from the forward shock interaction continuously heats the newly condensed dust in the same way that it heats the WMI component discussed in Section \ref{sec3:echo}.  Unlike the infrared echo scenario, however, the newly formed dust exists interior to the forward shock radius and the shock emission cannot be treated as a central point source.  

\subsection{A Multi-Component Model}

\begin{figure}
\begin{center}
\epsscale{1.3}
\plotone{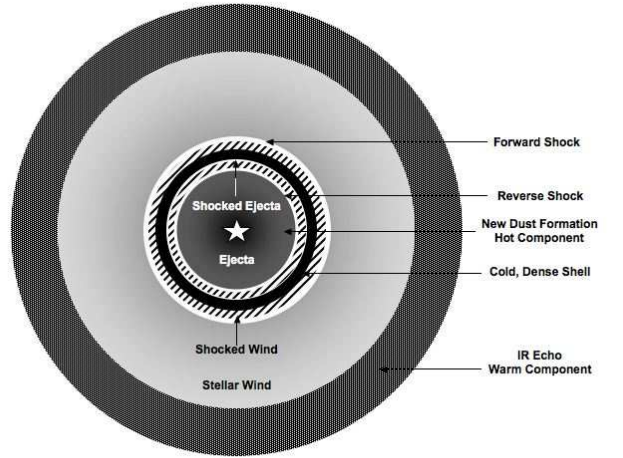}
\caption{Illustration of the proposed multi-component model for SN 2005ip consisting of both a hot, inner (HNI) component and warm, outer (WMI) component.  The HNI dust mass originates primarily from newly formed dust in the ejecta, while the WMI component likely originates from an circumstellar shock echo that forms from the heating of a large, pre-existing dust shell.  Collisional radiation from continuous shock interaction with the dense circumstellar medium generates the optical luminosity plateau \citep{smith09ip}, which is the likely heating mechanism for both the HNI and WMI components.
}
\label{f9}
\end{center}
\end{figure}

A multi-component dust model for SN 2005ip now begins to emerge, composed of both an inner, ``hot'' ($\sim$800 K), near-infrared (HNI) and an outer, ``warm'' ($\sim$400 K), mid-infrared (WMI) component.  Newly formed dust in the either the ejecta or cool, dense shell likely dominates the HNI component.  The WMI temperature and blackbody radius, $r_{bb} {\rm (WMI)} \approx $\radbbwarm, are consistent with a pre-existing dust shell heated by the observed late-time optical luminosity generated by the forward shock interaction (Section \ref{sec3:echo}).  Figure \ref{f9} illustrates the locations of each component, as well as the likely origins and heating mechanisms.

\begin{figure}
\begin{center}
\epsscale{1.2}
\plotone{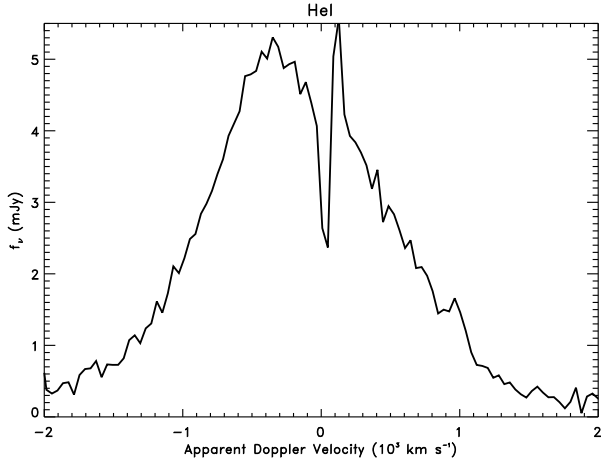}
\caption{The 1.083 \micron~He I line.  The shocked gas creates the broad ($\sim 1000$~\kms) emission feature.  The unshocked, slow moving, circumstellar medium formed by the progenitor wind produces the superimposed P Cygni profile.  The blue edge of the P Cygni absorption feature, $\sim$200 \kms, measures the wind velocity on day 862 post-discovery.}
\label{f10}
\end{center}
\end{figure}

The large pre-existing dust mass that contributes to the WMI flux component suggests significant mass loss from the progenitor.  Assuming a dust-to-gas ratio $Z_d = 0.01$, the observed WMI dust mass listed in Table \ref{tab2} ($M_d  {\rm (WMI)} \sim 0.01 - 0.05$~\msolar) yields a total gas mass of $M_g {\rm (WMI)} \sim $ 1 - 5 \msolar.  This mass accounts for the entire WMI emission, as the entire shell contributes to the observed flux given the size of the shell ($r_{bb} {\rm (WMI)} \approx $\radbbwarm) is significantly less than the observational time-scale (936 days).  The associated mass loss rate is 
\begin{eqnarray}
\mdot & = & \frac{M_g {\rm (WMI)}}{\Delta r} v_w\\ 
& = & 7.5 \times 10^{-3} \Big(\frac{M_g ({\rm WMI})}{\rm M_{\odot}}\Big) \times \\ 
& & \Big(\frac{v_w}{120~\rm km~s^{-1}}\Big) \Big(\frac{0.05~\rm ly}{r}\Big) \Big(\frac{r}{\Delta r}\Big) {\rm (M_{\odot}~yr^{-1})}. \nonumber
\end{eqnarray}
On day 413 post-discovery, \citet{smith09ip} measured a progenitor wind velocity $v_w = 120$~\kms.  Assuming a constant wind velocity and thin shell ($\frac{\Delta r}{r} = \frac{1}{10}$), the mass loss rate is $\mdot \approx 7.5 \times 10^{-2} - 3.8 \times 10^{-1} $ \ml, which is two-three orders of magnitude larger than calculated by \citet{smith09ip} at inner radii.   Furthermore, this mass loss rate is likely a lower limit.  He I P-Cygni profiles at 1.083 \micron~from the TripleSpec spectra show the progenitor wind velocity may have been closer to 200 \kms~at the WMI radius (Figure \ref{f10}), assuming the P-Cygni feature is generated by the cool, low velocity circumstellar environment that coincides with the WMI shell.  

The larger mass loss rate (and possibly faster progenitor wind velocity) of the WMI shell suggests a denser progenitor wind (or ``eruption'') occurred at $t \lesssim \frac{r_{bb} {\rm (WMI)}}{v_w} \approx \frac{4.8 \times 10^{11} {\rm~km}}{120 {\rm~km~s^{-1}}} \approx 125$~years prior to the core collapse.  Although \citet{smith09ip} conclude that the SN 2005ip progenitor was likely a red supergiant, these stars typically only have wind speeds $v_w \sim 20 - 40$~\kms~and mass loss rates up to $\dot{M} = 10^{-4} - 10^{-3}$ \ml~\citep{smith09rsg}.  The observed characteristics associated with the WMI shell are more consistent with Luminous Blue Variable (LBV) stars \citep[e.g.,][]{davidson89,humphreys94}, which can have wind speeds on order of hundreds \kms~\citep[e.g.,][]{leitherer97,kotakvink06} and can have mass loss rates up to $\dot{M} = 10^{-1}$ \ml~\citep{smith06a, smith06b,smith07}.

\section{Conclusion}
\label{sec3:con}

The $Spitzer$~spectrum presented here confirms the presence of warm dust in SN 2005ip.  Combined with near-infrared observations, the results show evidence for two independent dust masses: a hot, near-infrared (HNI) and warm, mid-infrared (WMI) component.  Infrared observations span the peak of the thermal emission, thereby providing strong constraints on the dust mass, temperature, and luminosity, which serve as critical diagnostics for disentangling the origin and heating mechanism of each component.  The HNI dust mass originates primarily from newly formed dust in the ejecta, or possibly the cool, dense shell, while the WMI component likely originates from an circumstellar shock echo that forms from the heating of a large, pre-existing dust shell.  For both components, the heating mechanism is likely the optical luminosity generated from the forward shock interaction with the circumstellar medium.  For wind speeds of $\sim 100 - 200$~\kms, the WMI dust shell likely formed via an eruption $\sim100$ years before the supernova.  These characteristics are consistent with a LBV progenitor, which has been linked to some core-collapse supernovae \citep{kotakvink06}, but is an emerging trend particularly within the Type IIn subclass \citep[e.g.,][]{gal-yam07,smith07, trundle08, smith08tf, smith09gy, gal-yam09}.

The newly formed HNI dust mass, $M_d (\rm HNI) \sim 5 \times 10^{-4}~$\msolar, is at least two orders of magnitude lower than predicted by ejecta condensation models to reproduce the large amounts of dust observed at high redshifts \citep{todini01,nozawa03,nozawa08}.  A dust mass of $\sim 10^{-4}~$\msolar~is comparable to other Type IIn events \citep[e.g.,][]{gerardy02, pozzo04, meikle07}.  For clumpy ejecta, it should be noted that both semi-analytical models \citep{varosi99} and Monte Carlo radiative-transfer simulations \citep{ercolano07} have shown that dust masses can be at least an order of magnitude larger than that predicted by the analysis presented in Section \ref{sec3:dust}.  The clump structure for any supernova, however, remains unconstrained at present.  The pre-existing WMI dust mass, $M_d (\rm WMI) \sim 0.05$~\msolar, is quite significant.  Although, in the case of SN 2005ip, the forward shock may ultimately destroy this dust (mid-infrared observations at later epochs will reveal the accuracy of this model), the forward shocks may be sufficiently decelerated in other systems to allow for dust survival.  

Late-time mid-infrared observations of dust in the supernova environment provide for a unique interpretation of the circumstellar environment and progenitor system.  Nonetheless, mid-infrared observations are quite rare.  Presented in this paper is the first mid-infrared spectrum of any Type IIn supernova.  In the future, we hope to grow the database of mid-infrared observations of Type IIn supernovae.  Doing so, however, can be slow as Type IIn events are particularly rare, consisting of only $\sim$2-3\% of all core-collapse supernovae \citep{gal-yam07} and occurring at a rate of no more than 10/yr out to 150 Mpc \citep{dahlen99}.  We therefore plan to revisit all Type IIn supernovae from the past ten years to determine the degree to which this subclass exhibits late-time dust emission and identify the emission mechanism.

\vspace{10 mm}

This work is based on observations made with the Spitzer Space Telescope (PID 50256), which is operated by the Jet Propulsion Laboratory, California Institute of Technology under a contract with NASA. Support for this work was provided by NASA through an award issued by JPL/Caltech.  O. D. F. is grateful for support from NASA GSRP, ARCS, and VSGC.  R. A. C. was supported by NSF grant AST-0807727.  The authors would like to thank Genevieve de Messieres, JD Smith, and the entire Spitzer Science Help Desk for their extensive help with the $Spitzer$~data reduction.  We are also grateful to Joel Sop for his help with the data reduction in the early stages.  We are also thankful to Sarah Doverspike and the anonymous referee for their helpful edits. 

\clearpage
\bibliographystyle{apj}
\bibliography{references}

\begin{thebibliography}{61}
\expandafter\ifx\csname natexlab\endcsname\relax\def\natexlab#1{#1}\fi

\bibitem[{Andrews {et~al.}(2010)Andrews, Gallagher, Clayton, Sugerman,
  Chatelain, Clem, Welch, Barlow, Ercolano, Fabbri, Wesson, \&
  Meixner}]{andrews10}
Andrews, J.~E., {et~al.} 2010, arXiv, 1004, 1209

\bibitem[{Bode \& Evans(1980)}]{bode80}
Bode, M.~F., \& Evans, A. 1980, A{\&}A, 89, 158

\bibitem[{Boles {et~al.}(2005)Boles, Nakano, \& Itagaki}]{boles05}
Boles, T., Nakano, S., \& Itagaki, K. 2005, IAU Circ., 275, 1

\bibitem[{Chugai \& Danziger(1994)}]{chugai94}
Chugai, N.~N., \& Danziger, I.~J. 1994, \mnras, 268, 173

\bibitem[{Cushing {et~al.}(2004)Cushing, Vacca, \& Rayner}]{cushing04}
Cushing, M.~C., Vacca, W.~D., \& Rayner, J.~T. 2004, \pasp, 116, 362

\bibitem[{Dahl{\'e}n \& Fransson(1999)}]{dahlen99}
Dahl{\'e}n, T., \& Fransson, C. 1999, A{\&}A, 350, 349

\bibitem[{Davidson(1989)}]{davidson89}
Davidson, K. 1989, in Physics of Luminous Blue Variables, eds. K. Davidson, A.
  F. J. Moffat, \& H. J. G. L. M. Lamers, Proc. of IAU Colloq. 113, Vol. 157,
  101

\bibitem[{Draine \& Li(2001)}]{draine01}
Draine, B.~T., \& Li, A. 2001, \apj, 551, 807

\bibitem[{Dwek(1983)}]{dwek83}
Dwek, E. 1983, \apj, 274, 175

\bibitem[{{Dwek}(1987)}]{dwek87}
{Dwek}, E. 1987, \apj, 322, 812

\bibitem[{Dwek \& Arendt(1992)}]{dwek92}
Dwek, E., \& Arendt, R.~G. 1992, ARA\&A, 30, 11

\bibitem[{Dwek {et~al.}(2008)Dwek, Arendt, Bouchet, Burrows, Challis, Danziger,
  Buizer, Gehrz, Kirshner, McCray, Park, Polomski, \& Woodward}]{dwek08}
Dwek, E., {et~al.} 2008, \apj, 676, 1029

\bibitem[{Elmhamdi {et~al.}(2004)Elmhamdi, Danziger, Cappellaro, Valle,
  Gouiffes, Phillips, \& Turatto}]{elmhamdi04}
Elmhamdi, A., Danziger, I.~J., Cappellaro, E., Valle, M.~D., Gouiffes, C.,
  Phillips, M.~M., \& Turatto, M. 2004, A{\&}A, 426, 963

\bibitem[{Elmhamdi {et~al.}(2003)Elmhamdi, Danziger, Chugai, Pastorello,
  Turatto, Cappellaro, Altavilla, Benetti, Patat, \& Salvo}]{elmhamdi03}
Elmhamdi, A., {et~al.} 2003, MNRAS, 338, 939

\bibitem[{Ercolano {et~al.}(2007)Ercolano, Barlow, \& Sugerman}]{ercolano07}
Ercolano, B., Barlow, M.~J., \& Sugerman, B. E.~K. 2007, MNRAS, 375, 753

\bibitem[{Fazio {et~al.}(2004)Fazio, Hora, Allen, Ashby, Barmby, Deutsch,
  Huang, Kleiner, Marengo, Megeath, Melnick, Pahre, Patten, Polizotti, Smith,
  Taylor, Wang, Willner, Hoffmann, Pipher, Forrest, McMurty, McCreight,
  McKelvey, McMurray, Koch, Moseley, Arendt, Mentzell, Marx, Losch, Mayman,
  Eichhorn, Krebs, Jhabvala, Gezari, Fixsen, Flores, Shakoorzadeh, Jungo,
  Hakun, Workman, Karpati, Kichak, Whitley, Mann, Tollestrup, Eisenhardt,
  Stern, Gorjian, Bhattacharya, Carey, Nelson, Glaccum, Lacy, Lowrance, Laine,
  Reach, Stauffer, Surace, Wilson, Wright, Hoffman, Domingo, \&
  Cohen}]{fazio04}
Fazio, G.~G., {et~al.} 2004, ApJS, 154, 10

\bibitem[{Fox {et~al.}(2009)Fox, Skrutskie, Chevalier, Kanneganti, Park,
  Wilson, Nelson, Amirhadji, Crump, Hoeft, Provence, Sargeant, Sop, Tea,
  Thomas, \& Woolard}]{fox09}
Fox, O., {et~al.} 2009, \apj, 691, 650

\bibitem[{Gal-Yam \& Leonard(2009)}]{gal-yam09}
Gal-Yam, A., \& Leonard, D.~C. 2009, Nature, 458, 865

\bibitem[{Gal-Yam {et~al.}(2007)Gal-Yam, Leonard, Fox, Cenko, Soderberg, Moon,
  Sand, Li, Filippenko, Aldering, \& Copin}]{gal-yam07}
Gal-Yam, A., {et~al.} 2007, \apj, 656, 372

\bibitem[{Gerardy {et~al.}(2002)Gerardy, Fesen, Nomoto, Garnavich, Jha,
  Challis, Kirshner, H{\"o}flich, \& Wheeler}]{gerardy02}
Gerardy, C.~L., {et~al.} 2002, \apj, 575, 1007

\bibitem[{Herter {et~al.}(2008)Herter, Henderson, Wilson, Matthews, Rahmer,
  Bonati, Muirhead, Adams, Lloyd, Skrutskie, Moon, Parshley, Nelson,
  Martinache, \& Gull}]{herter08}
Herter, T.~L., {et~al.} 2008, Proc. of SPIE, 7014, 30

\bibitem[{Higdon {et~al.}(2004)Higdon, Devost, Higdon, Brandl, Houck, Hall,
  Barry, Charmandaris, Smith, Sloan, \& Green}]{higdon04}
Higdon, S. J.~U., {et~al.} 2004, \pasp, 116, 975

\bibitem[{Houck {et~al.}(2004)Houck, Roellig, van Cleve, Forrest, Herter,
  Lawrence, Matthews, Reitsema, Soifer, Watson, Weedman, Huisjen, Troeltzsch,
  Barry, Bernard-Salas, Blacken, Brandl, Charmandaris, Devost, Gull, Hall,
  Henderson, Higdon, Pirger, Schoenwald, Sloan, Uchida, Appleton, Armus,
  Burgdorf, Fajardo-Acosta, Grillmair, Ingalls, Morris, \& Teplitz}]{houck04}
Houck, J.~R., {et~al.} 2004, ApJS, 154, 18

\bibitem[{Humphreys \& Davidson(1994)}]{humphreys94}
Humphreys, R.~M., \& Davidson, K. 1994, PASP, 106, 1025

\bibitem[{Kotak(2008)}]{kotak08}
Kotak, R. 2008, Proc. of the IAU Symposium, 250, 437

\bibitem[{Kotak \& Vink(2006)}]{kotakvink06}
Kotak, R., \& Vink, J.~S. 2006, A{\&}A, 460, L5

\bibitem[{Kotak {et~al.}(2009)Kotak, Meikle, Farrah, Gerardy, Foley, Dyk,
  Fransson, Lundqvist, Sollerman, Fesen, Filippenko, Mattila, Silverman,
  Andersen, H{\"o}flich, Pozzo, \& Wheeler}]{kotak09}
Kotak, R., {et~al.} 2009, \apj, 704, 306

\bibitem[{Lebouteiller {et~al.}(2009)Lebouteiller, Bernard-Salas, Sloan, \&
  Barry}]{lebouteiller09}
Lebouteiller, V., Bernard-Salas, J., Sloan, G.~C., \& Barry, D.~J. 2009, PASP,
  122, 231

\bibitem[{Leitherer(1997)}]{leitherer97}
Leitherer, C. 1997, in Luminous Blue Variables: Massive Stars in Transition,
  eds. Nota and Lamers, ASP Conf., Vol. 120, 58

\bibitem[{Lucy {et~al.}(1991)Lucy, Danziger, Gouiffes, \& Bouchet}]{lucy91}
Lucy, L.~B., Danziger, I.~J., Gouiffes, C., \& Bouchet, P. 1991, in Supernovae,
  ed. S.~E. Woosley (New York: Springer), 82

\bibitem[{Mattila {et~al.}(2008)Mattila, Meikle, Lundqvist, Pastorello, Kotak,
  Eldridge, Smartt, Adamson, Gerardy, Rizzi, Stephens, \& Dyk}]{mattila08}
Mattila, S., {et~al.} 2008, MNRAS, 389, 141

\bibitem[{Meikle {et~al.}(1993)Meikle, Spyromilio, Allen, Varani, \&
  Cumming}]{meikle93}
Meikle, W. P.~S., Spyromilio, J., Allen, D.~A., Varani, G.-F., \& Cumming,
  R.~J. 1993, MNRAS, 261, 535

\bibitem[{{Meikle} {et~al.}(2007){Meikle}, {Mattila}, {Pastorello}, {Gerardy},
  {Kotak}, {Sollerman}, {Van Dyk}, {Farrah}, {Filippenko}, {H{\"o}flich},
  {Lundqvist}, {Pozzo}, \& {Wheeler}}]{meikle07}
{Meikle}, W.~P.~S., {et~al.} 2007, \apj, 665, 608

\bibitem[{Miller {et~al.}(2010{\natexlab{a}})Miller, Smith, Li, Bloom,
  Chornock, Filippenko, \& Prochaska}]{miller10gy}
Miller, A.~A., Smith, N., Li, W., Bloom, J.~S., Chornock, R., Filippenko,
  A.~V., \& Prochaska, J.~X. 2010{\natexlab{a}}, \aj, 139, 2218

\bibitem[{Miller {et~al.}(2010{\natexlab{b}})Miller, Silverman, Butler, Bloom,
  Chornock, Filippenko, Ganeshalingam, Klein, Li, Nugent, Smith, \&
  Steele}]{miller10}
Miller, A.~A., {et~al.} 2010{\natexlab{b}}, MNRAS, 404, 305

\bibitem[{Modjaz {et~al.}(2005)Modjaz, Kirshner, Challis, \&
  Calkins}]{modjaz05}
Modjaz, M., Kirshner, R., Challis, P., \& Calkins, M. 2005, IAU Circ., 8628, 2

\bibitem[{{Nozawa} {et~al.}(2003){Nozawa}, {Kozasa}, {Umeda}, {Maeda}, \&
  {Nomoto}}]{nozawa03}
{Nozawa}, T., {Kozasa}, T., {Umeda}, H., {Maeda}, K., \& {Nomoto}, K. 2003,
  \apj, 598, 785

\bibitem[{{Nozawa} {et~al.}(2008){Nozawa}, {Kozasa}, {Tominaga}, {Sakon},
  {Tanaka}, {Suzuki}, {Nomoto}, {Maeda}, {Umeda}, {Limongi}, \&
  {Onaka}}]{nozawa08}
{Nozawa}, T., {et~al.} 2008, \apj, 684, 1343

\bibitem[{Pastorello {et~al.}(2002)Pastorello, Turatto, Benetti, Cappellaro,
  Danziger, Mazzali, Patat, Filippenko, Schlegel, \& Matheson}]{pastorello02}
Pastorello, A., {et~al.} 2002, MNRAS, 333, 27

\bibitem[{Pozzo {et~al.}(2004)Pozzo, Meikle, Fassia, Geballe, Lundqvist,
  Chugai, \& Sollerman}]{pozzo04}
Pozzo, M., Meikle, W. P.~S., Fassia, A., Geballe, T., Lundqvist, P., Chugai,
  N.~N., \& Sollerman, J. 2004, MNRAS, 352, 457

\bibitem[{Quimby {et~al.}(2007)Quimby, Aldering, Wheeler, H{\"o}flich, Akerlof,
  \& Rykoff}]{quimby07}
Quimby, R.~M., Aldering, G., Wheeler, J.~C., H{\"o}flich, P., Akerlof, C.~W.,
  \& Rykoff, E.~S. 2007, \apj, 668, L99

\bibitem[{Rest {et~al.}(2009)Rest, Foley, Gezari, Narayan, Draine, Olsen,
  Huber, Matheson, Garg, Welch, Becker, Challis, Clocchiatti, Cook, Damke,
  Meixner, Miknaitis, Minniti, Morelli, Nikolaev, Pignata, Prieto, Smith,
  Stubbs, Suntzeff, Walker, Wood-Vasey, Zenteno, Wyrzykowski, Udalski,
  Szymanski, Kubiak, Pietrzynski, Soszynski, Szewczyk, Ulaczyk, \&
  Poleski}]{rest09}
Rest, A., {et~al.} 2009, arXiv:0911.2002

\bibitem[{Salamanca {et~al.}(2002)Salamanca, Terlevich, \&
  Tenorio-Tagle}]{salamanca02}
Salamanca, I., Terlevich, R.~J., \& Tenorio-Tagle, G. 2002, MNRAS, 330, 844

\bibitem[{Schlegel(1990)}]{schlegel90}
Schlegel, E.~M. 1990, MNRAS, 244, 269

\bibitem[{Smith {et~al.}(2008{\natexlab{a}})Smith, Chornock, Li, Ganeshalingam,
  Silverman, Foley, Filippenko, \& Barth}]{smith08tf}
Smith, N., Chornock, R., Li, W., Ganeshalingam, M., Silverman, J.~M., Foley,
  R.~J., Filippenko, A.~V., \& Barth, A.~J. 2008{\natexlab{a}}, \apj, 686, 467

\bibitem[{Smith {et~al.}(2010)Smith, Chornock, Silverman, Filippenko, \&
  Foley}]{smith09gy}
Smith, N., Chornock, R., Silverman, J.~M., Filippenko, A.~V., \& Foley, R.~J.
  2010, \apj, 709, 856

\bibitem[{Smith {et~al.}(2008{\natexlab{b}})Smith, Foley, \&
  Filippenko}]{smith08jc}
Smith, N., Foley, R.~J., \& Filippenko, A.~V. 2008{\natexlab{b}}, \apj, 680,
  568

\bibitem[{Smith \& Hartigan(2006)}]{smith06b}
Smith, N., \& Hartigan, P. 2006, \apj, 638, 1045

\bibitem[{Smith {et~al.}(2009{\natexlab{a}})Smith, Hinkle, \&
  Ryde}]{smith09rsg}
Smith, N., Hinkle, K.~H., \& Ryde, N. 2009{\natexlab{a}}, \aj, 137, 3558

\bibitem[{Smith \& Owocki(2006)}]{smith06a}
Smith, N., \& Owocki, S.~P. 2006, \apj, 645, L45

\bibitem[{Smith {et~al.}(2007)Smith, Li, Foley, Wheeler, Pooley, Chornock,
  Filippenko, Silverman, Quimby, Bloom, \& Hansen}]{smith07}
Smith, N., {et~al.} 2007, \apj, 666, 1116

\bibitem[{Smith {et~al.}(2009{\natexlab{b}})Smith, Silverman, Chornock,
  Filippenko, Wang, Li, Ganeshalingam, Foley, Rex, \& Steele}]{smith09ip}
---. 2009{\natexlab{b}}, \apj, 695, 1334

\bibitem[{Soderberg {et~al.}(2008)Soderberg, Berger, Page, Schady, Parrent,
  Pooley, Wang, Ofek, Cucchiara, Rau, Waxman, Simon, Bock, Milne, Page,
  Barentine, Barthelmy, Beardmore, Bietenholz, Brown, Burrows, Burrows,
  Byrngelson, Cenko, Chandra, Cummings, Fox, Gal-Yam, Gehrels, Immler,
  Kasliwal, Kong, Krimm, Kulkarni, Maccarone, M{\'e}sz{\'a}ros, Nakar, O'Brien,
  Overzier, Pasquale, Racusin, Rea, \& York}]{soderberg08}
Soderberg, A.~M., {et~al.} 2008, Nature, 453, 469

\bibitem[{Steele {et~al.}(2008)Steele, Ganeshalingam, Chornock, \&
  Filippenko}]{green08}
Steele, T.~N., Ganeshalingam, M., Chornock, R., \& Filippenko, A.~V. 2008, IAU
  Circ., 1465, 4

\bibitem[{Sugerman {et~al.}(2006)Sugerman, Ercolano, Barlow, Tielens, Clayton,
  Zijlstra, Meixner, Speck, Gledhill, Panagia, Cohen, Gordon, Meyer, Fabbri,
  Bowey, Welch, Regan, \& Kennicutt}]{sugerman06}
Sugerman, B. E.~K., {et~al.} 2006, Science, 313, 196

\bibitem[{Todini \& Ferrara(2001)}]{todini01}
Todini, P., \& Ferrara, A. 2001, MNRAS, 325, 726

\bibitem[{Trundle {et~al.}(2008)Trundle, Kotak, Vink, \& Meikle}]{trundle08}
Trundle, C., Kotak, R., Vink, J.~S., \& Meikle, W. P.~S. 2008, A{\&}A, 483, L47

\bibitem[{Usov(1991)}]{usov91}
Usov, V.~V. 1991, MNRAS, 252, 49

\bibitem[{V{\'a}rosi \& Dwek(1999)}]{varosi99}
V{\'a}rosi, F., \& Dwek, E. 1999, \apj, 523, 265

\bibitem[{Williams {et~al.}(2006)Williams, Borkowski, Reynolds, Blair,
  Ghavamian, Hendrick, Long, Points, Raymond, Sankrit, Smith, \&
  Winkler}]{williams06}
Williams, B.~J., {et~al.} 2006, \apj, 652, L33

\bibitem[{Wilson {et~al.}(2004)Wilson, Henderson, Herter, Matthews, Skrutskie,
  Adams, Moon, Smith, Gautier, Ressler, Soifer, Lin, Howard, LaMarr, Stolberg,
  \& Zink}]{wilson04}
Wilson, J.~C., {et~al.} 2004, Proc. of SPIE, 5492, 1295

\end{thebibliography}

\end{document}